\documentclass[conference]{IEEEtran}
\IEEEoverridecommandlockouts
\usepackage{cite}
\usepackage{amsmath,amssymb,amsfonts}
\usepackage{graphicx}
\usepackage{booktabs}
\usepackage{multirow}
\usepackage{makecell}
\usepackage[inline]{enumitem}
\usepackage{comment}
\usepackage{textcomp}
\usepackage[dvipsnames]{xcolor}
\usepackage{listing}
\usepackage{fvextra}
\input{pygments-style.tex}
\usepackage{xurl}
\usepackage[hidelinks]{hyperref}
\usepackage{tikz}
\usetikzlibrary{tikzmark, fit, calc, arrows.meta}

\usepackage[english, status=draft, margin=false, inline=true]{fixme}
\fxusetheme{color}
\FXRegisterAuthor{jf}{ajf}{\color{red}JF}
\newcommand{\rmspace}{}

\newcommand{\std}[1]{%
  \textbf{{\color{gray}\fontsize{7}{8.4}\selectfont(\textpm{} #1)}}%
}

\begin{document}

\title{STRATOS: Bridging the Symbolic-to-Numeric Gap\\in Spatio-Temporal Text-to-SQL\\for Meteorological Data
}

\author{\IEEEauthorblockN{Yi Zhang\IEEEauthorrefmark{1},
Farhad Nooralahzadeh\IEEEauthorrefmark{1},
Jonathan Fürst\IEEEauthorrefmark{1},
Fabio Scherrer\IEEEauthorrefmark{1},\\
Claude Lehmann\IEEEauthorrefmark{1},
Antonis Bezes\IEEEauthorrefmark{2},
Vassiliki Kotroni\IEEEauthorrefmark{2},
Kurt Stockinger\IEEEauthorrefmark{1}}
\IEEEauthorblockA{{\small\IEEEauthorrefmark{1}\textit{Zurich University of Applied Sciences, Switzerland;}
\IEEEauthorrefmark{2}\textit{National Observatory of Athens, Greece;}}\\
\small\ \texttt{yi.zhang@zhaw.ch; farhad.nooralahzadeh@zhaw.ch; jonathan.fuerst@zhaw.ch;}\\
\small\ \texttt{fabio.scherrer@zhaw.ch; claude.lehmann@zhaw.ch; antbezes@noa.gr;}\\
\small\ \texttt{kotroni@noa.gr; kurt.stockinger@zhaw.ch}}}

\maketitle

\begin{abstract}
Copernicus, the European Union's Earth observation program, produces petabytes of Earth observation and climate data, offering immense potential for research, policy, and applications. However, access to these datasets requires advanced programming skills and familiarity with domain-specific formats such as NetCDF or GRIB. Moreover, general-purpose Text-to-SQL systems fail when applied naively to the meteorological domain due to a profound ``Symbolic-to-Numeric'' gap.
To overcome these limitations, we present an end-to-end Text-to-SQL framework specifically engineered for real-world, scalable meteorological data exploration. Our system intercepts natural language to resolve spatial and semantic ambiguities \textit{before} SQL generation. We design STRATOS, a Spatio-Temporal Resolution Agent for Text-to-SQL to dynamically bridge the symbolic-to-numeric gap by mapping fuzzy concepts to a localized ontology and resolving spatial entities via external knowledge bases. Further, our complexity-aware query rewriter rewrites expensive spatial predicates, reducing the mean query execution time by a factor of 2 and avoiding timeout queries. Last, we introduce the STRATOS Evaluation Workload, comprising 7,520 complex query pairs explicitly designed by domain experts to test scalability and symbolic-to-numeric translation across challenging spatio-temporal dimensions previously unexplored by Text-to-SQL systems.
\end{abstract}


\section{Introduction}

Copernicus, the European Union's Earth observation program, produces petabytes of Earth observation and climate data, with immense potential for research, policy, and applications. However, there is a major obstacle: access to these datasets requires advanced programming skills and familiarity with fragmented, domain-specific formats such as NetCDF~\cite{rew1990netcdf} or GRIB~\cite{grib2}. Even for domain experts, the ``time-to-insight'' is throttled by the need to manually align disparate sources, such as the ERA5-Land~\cite{munoz2021era5} atmospheric reanalysis ($0.1^\circ \times 0.1^\circ$ resolution) and CORINE Land Cover~\cite{clc2018_raster} maps (100 m resolution).

While migrating this multidimensional scientific data into a unified relational database architecture is a foundational prerequisite for declarative querying, it is not a complete solution. General-purpose Large Language Models (LLMs) and Text-to-SQL systems appear perfectly suited to bridge the remaining usability gap, yet they fail when applied naively to the meteorological domain. This failure is rooted in a profound semantic and structural disconnect.

Consider the seemingly simple query: \textit{``How many heat waves did Athens experience in the first two decades of the 21st century?''} Answering this question is challenging because the definition of a heat wave varies by local climatology, available variables, and persistence criteria~\cite{galanaki2022spatio}. Considering a working definition of a daily maximum temperature ($T_{max}$) $\ge 35^\circ\text{C}$ for at least three consecutive days, generating a correct SQL query (Figure~\ref{fig:meteo_problem}) exposes three distinct failure points for standard Text-to-SQL systems:

\begin{figure*}[t]
  \centering
  \newcommand{\lbl}[1]{\textcolor{BrickRed}{\sffamily #1}}
\begin{Verbatim}[commandchars=\\\{\},codes={\catcode`\$=3\catcode`\^=7\catcode`\_=8\relax},fontsize=\footnotesize,frame=lines,framesep=2mm]
\PYG{k}{WITH}\PYG{+w}{ }\PYG{n}{filtered}\PYG{+w}{ }\PYG{k}{AS}\PYG{+w}{ }\PYG{p}{(}
\PYG{+w}{  }\PYG{k}{SELECT}\PYG{+w}{ }\PYG{k}{time}\PYG{p}{,}\PYG{+w}{ }\PYG{n}{tmax}\PYG{p}{,}\PYG{+w}{ }\PYG{n+nb}{DATE}\PYG{p}{(}\PYG{k}{time}\PYG{p}{)}\PYG{+w}{ }\PYG{k}{AS}\PYG{+w}{ }\PYG{n}{date\PYGZus{}only}
\PYG{+w}{  }\PYG{k}{FROM}\PYG{+w}{ }\PYG{n}{meteo\PYGZus{}tmax}
\PYG{+w}{  }\PYG{k}{WHERE}\PYG{+w}{ }\PYG{k}{EXTRACT}\PYG{p}{(}\PYG{k}{YEAR}\PYG{+w}{ }\PYG{k}{FROM}\PYG{+w}{ }\PYG{k}{time}\PYG{p}{)}\PYG{+w}{ }\PYG{k}{BETWEEN}\PYG{+w}{ }\PYG{l+m+mi}{2001}\PYG{+w}{ }\PYG{k}{AND}\PYG{+w}{ }\PYG{l+m+mi}{2020}
\PYG{+w}{    }\PYG{esc}{\tikzmark{s1}}\PYG{k}{AND}\PYG{+w}{ }\PYG{n}{round}\PYG{p}{(}\PYG{n}{latitude}\PYG{p}{:}\PYG{p}{:}\PYG{n+nb}{numeric}\PYG{p}{,}\PYG{+w}{ }\PYG{l+m+mi}{1}\PYG{p}{)}\PYG{+w}{ }\PYG{o}{=}\PYG{+w}{ }\PYG{l+m+mi}{38}\PYG{+w}{                          }\PYG{esc}{\tikzmark{l1}\lbl{Coordinates of Athens, Greece}}
\PYG{+w}{    }\PYG{k}{AND}\PYG{+w}{ }\PYG{n}{round}\PYG{p}{(}\PYG{n}{longitude}\PYG{p}{:}\PYG{p}{:}\PYG{n+nb}{numeric}\PYG{p}{,}\PYG{+w}{ }\PYG{l+m+mi}{1}\PYG{p}{)}\PYG{+w}{ }\PYG{o}{=}\PYG{+w}{ }\PYG{l+m+mi}{23}\PYG{p}{.}\PYG{l+m+mi}{7}\PYG{esc}{\tikzmark{e1}}
\PYG{+w}{    }\PYG{k}{AND}\PYG{+w}{ }\PYG{n}{tmax}\PYG{+w}{ }\PYG{k}{IS}\PYG{+w}{ }\PYG{k}{NOT}\PYG{+w}{ }\PYG{k}{NULL}\PYG{+w}{ }\PYG{p}{)}\PYG{p}{,}
\PYG{n}{heatwave\PYGZus{}days}\PYG{+w}{ }\PYG{k}{AS}\PYG{+w}{ }\PYG{p}{(}
\PYG{+w}{  }\PYG{k}{SELECT}\PYG{+w}{ }\PYG{n}{date\PYGZus{}only}\PYG{p}{,}\PYG{+w}{ }\PYG{n}{tmax}\PYG{p}{,}\PYG{+w}{                                            }\PYG{esc}{\tikzmark{l2}\lbl{Kelvin to Celsius conversion}}
\PYG{+w}{         }\PYG{esc}{\tikzmark{s3}}\PYG{k}{CASE}\PYG{+w}{ }\PYG{k}{WHEN}\PYG{+w}{ }\PYG{esc}{\tikzmark{s2}}\PYG{n}{tmax}\PYG{+w}{ }\PYG{o}{\PYGZhy{}}\PYG{+w}{ }\PYG{l+m+mi}{273}\PYG{p}{.}\PYG{l+m+mi}{15}\PYG{esc}{\tikzmark{e2}}\PYG{+w}{ }\PYG{o}{\PYGZgt{}}\PYG{o}{=}\PYG{+w}{ }\PYG{l+m+mi}{35}\PYG{p}{.}\PYG{l+m+mi}{0}\PYG{+w}{ }\PYG{k}{THEN}\PYG{+w}{ }\PYG{l+m+mi}{1}\PYG{+w}{ }\PYG{k}{ELSE}\PYG{+w}{ }\PYG{l+m+mi}{0}\PYG{+w}{ }\PYG{k}{END}\PYG{esc}{\tikzmark{e3}}\PYG{+w}{          }\PYG{esc}{\tikzmark{l3}\lbl{Definition of a heat day}}
\PYG{+w}{           }\PYG{k}{AS}\PYG{+w}{ }\PYG{n}{is\PYGZus{}heatwave\PYGZus{}day}
\PYG{+w}{  }\PYG{k}{FROM}\PYG{+w}{ }\PYG{n}{filtered}\PYG{+w}{ }\PYG{p}{)}\PYG{p}{,}
\PYG{esc}{\tikzmark{s4}}\PYG{n}{streaks}\PYG{+w}{ }\PYG{k}{AS}\PYG{+w}{ }\PYG{p}{(}
\PYG{+w}{  }\PYG{k}{SELECT}\PYG{+w}{ }\PYG{n}{date\PYGZus{}only}\PYG{p}{,}\PYG{+w}{ }\PYG{n}{is\PYGZus{}heatwave\PYGZus{}day}\PYG{p}{,}
\PYG{+w}{         }\PYG{n}{date\PYGZus{}only}\PYG{+w}{ }\PYG{o}{\PYGZhy{}}\PYG{+w}{ }\PYG{p}{(}\PYG{n}{ROW\PYGZus{}NUMBER}\PYG{p}{(}\PYG{p}{)}\PYG{+w}{ }\PYG{n}{OVER}\PYG{+w}{ }\PYG{p}{(}\PYG{k}{ORDER}\PYG{+w}{ }\PYG{k}{BY}\PYG{+w}{ }\PYG{n}{date\PYGZus{}only}\PYG{p}{)}\PYG{esc}{\tikzmark{r4}}
\PYG{+w}{                      }\PYG{o}{*}\PYG{+w}{ }\PYG{n+nb}{INTERVAL}\PYG{+w}{ }\PYG{l+s+s1}{\PYGZsq{}1 day\PYGZsq{}}\PYG{p}{)}\PYG{+w}{ }\PYG{k}{AS}\PYG{+w}{ }\PYG{n}{group\PYGZus{}id}
\PYG{+w}{  }\PYG{k}{FROM}\PYG{+w}{ }\PYG{n}{heatwave\PYGZus{}days}
\PYG{+w}{  }\PYG{k}{WHERE}\PYG{+w}{ }\PYG{n}{is\PYGZus{}heatwave\PYGZus{}day}\PYG{+w}{ }\PYG{o}{=}\PYG{+w}{ }\PYG{l+m+mi}{1}\PYG{+w}{ }\PYG{p}{)}\PYG{esc}{\tikzmark{e4}}\PYG{p}{,}
\PYG{n}{streak\PYGZus{}lengths}\PYG{+w}{ }\PYG{k}{AS}\PYG{+w}{ }\PYG{p}{(}
\PYG{+w}{  }\PYG{k}{SELECT}\PYG{+w}{ }\PYG{n}{group\PYGZus{}id}\PYG{p}{,}\PYG{+w}{ }\PYG{k}{COUNT}\PYG{p}{(}\PYG{o}{*}\PYG{p}{)}\PYG{+w}{ }\PYG{k}{AS}\PYG{+w}{ }\PYG{n}{streak\PYGZus{}length}
\PYG{+w}{  }\PYG{k}{FROM}\PYG{+w}{ }\PYG{n}{streaks}
\PYG{+w}{  }\PYG{k}{GROUP}\PYG{+w}{ }\PYG{k}{BY}\PYG{+w}{ }\PYG{n}{group\PYGZus{}id}\PYG{+w}{                                                  }\PYG{esc}{\tikzmark{l4}\lbl{Definition of a heat wave}}
\PYG{+w}{  }\PYG{k}{HAVING}\PYG{+w}{ }\PYG{esc}{\tikzmark{s5}}\PYG{k}{COUNT}\PYG{p}{(}\PYG{o}{*}\PYG{p}{)}\PYG{+w}{ }\PYG{o}{\PYGZgt{}}\PYG{o}{=}\PYG{+w}{ }\PYG{l+m+mi}{3}\PYG{esc}{\tikzmark{e5}}\PYG{+w}{ }\PYG{p}{)}
\PYG{k}{SELECT}\PYG{+w}{ }\PYG{k}{COUNT}\PYG{p}{(}\PYG{o}{*}\PYG{p}{)}\PYG{+w}{ }\PYG{k}{AS}\PYG{+w}{ }\PYG{n}{heat\PYGZus{}wave\PYGZus{}count}
\PYG{k}{FROM}\PYG{+w}{ }\PYG{n}{streak\PYGZus{}lengths}\PYG{p}{;}
\end{Verbatim}
\begin{tikzpicture}[remember picture, overlay,
    box/.style={draw=BrickRed, thick, inner sep=1pt},
    arr/.style={-{Stealth[length=4pt]}, BrickRed, thick}]
  \node[box, fit={($(pic cs:s1)+(0,5.5pt)$) ($(pic cs:e1)+(0,-1.5pt)$)}] (b1) {};
  \node[box, dashed, inner sep=0.5pt, fit={($(pic cs:s2)+(0,5.5pt)$) ($(pic cs:e2)+(0,-1.5pt)$)}] (b2) {};
  \node[box, inner sep=2pt, fit={($(pic cs:s3)+(0,5.5pt)$) ($(pic cs:e3)+(0,-1.5pt)$)}] (b3) {};
  \node[box, fit={($(pic cs:s4)+(0,5.5pt)$) (pic cs:r4) ($(pic cs:e4)+(0,-1.5pt)$)}] (b4) {};
  \node[box, fit={($(pic cs:s5)+(0,5.5pt)$) ($(pic cs:e5)+(0,-1.5pt)$)}] (b5) {};
  \draw[arr] ($(pic cs:l1)+(-3pt,2.5pt)$) -- (b1.east);
  \draw[arr, dashed] ($(pic cs:l2)+(-3pt,2.5pt)$) -| ($(b2.north)!0.5!(b2.north east)$);
  \draw[arr] ($(pic cs:l3)+(-3pt,2.5pt)$) -- (b3.east);
  \draw[arr] ($(pic cs:l4)+(-3pt,4pt)$) -- (b4.south east);
  \draw[arr] ($(pic cs:l4)+(-3pt,1pt)$) -- (b5.east);
\end{tikzpicture}
  \caption{How many heat waves did Athens experience in the first two decades of the 21st century? Answering this question requires \emph{spatial decoding (coordinates of Athens, Greece)} and \emph{semantic translation (a definition of a heat day and heat wave with unit conversion from Kelvin to Celsius)}.}
  \label{fig:meteo_problem}
\end{figure*}

\begin{itemize}[leftmargin=0em]
    \item \textbf{Location Resolution (The ``Where''):} ``City of Athens'' is a symbolic toponym. Because the database stores values against specific latitude/longitude grids, the query requires precise mathematical geometries (e.g., a spatial bounding box or multipolygon). Relying on an LLM to guess coordinate boundaries yields hallucinations or omits the spatial filter entirely.
    
    \item \textbf{The Symbolic-to-Numeric Gap (The ``What''):} ``Heat waves'' is a derived meteorological index, not a database column. The system must map natural language concepts to formal threshold-duration definitions, applying the correct meteorological variables and implicit unit conversions (e.g., Kelvin to Celsius), which are invisible to a Text-to-SQL system.
    
    \item \textbf{Computational Complexity (The ``How''):} Even if a Text-to-SQL system generates a semantically correct SQL query, executing naive spatial predicates (e.g., massive \texttt{WHERE ... IN} coordinate lists or complex vector-to-raster joins) over hundreds of gigabytes of gridded array data can take hours, making interactive data exploration impossible.
\end{itemize}

To overcome these limitations, we present STRATOS, a Spatio-Temporal Resolution Agent for Text-to-SQL, specifically engineered for scalable, meteorological data science. Our system intercepts natural language to resolve spatial and semantic ambiguities \textit{before} SQL generation and employs deterministic query rewriting to optimize execution. Our core contributions are threefold:

\begin{enumerate}[leftmargin=1em, label=\textbf{\arabic*.}]
    \item \textbf{Spatio-Temporal Intent Resolution:} Unlike standard Text-to-SQL systems that rely on purely schema-bound semantic matching, we propose an externally-augmented pipeline to dynamically bridge the symbolic-to-numeric gap. We extract and resolve location entities against external knowledge bases (Wikidata/OpenStreetMap) to inject precise spatial bounds. Concurrently, a Domain Logic Injector maps fuzzy meteorological concepts to a localized ontology, ensuring scientific reproducibility and preventing LLM hallucination.
    
    \item \textbf{Complexity-Aware Spatial Run-time Rewriting:} While generating a structurally correct SQL query is becoming increasingly easy for modern Text-to-SQL systems~\cite{zhu2026revisql, li2024can}, generating an \textit{efficient} query over hundreds of gigabytes of data remains a challenge. We bypass the slow execution of naive Text-to-SQL outputs by dynamically synthesizing high-performance spatial SQL. By rewriting expensive \texttt{IN}-list clauses into \texttt{VALUES}-based subqueries and leveraging spatial expression indices, we reduce the mean query execution time by a factor of 2.3 and avoid all timeouts.
    
    \item \textbf{A Scalable Evaluation Workload:} We introduce the STRATOS Evaluation Workload, explicitly designed to test scalability and symbolic-to-numeric translation. Based on 161 manually curated, domain-expert-verified seed queries spanning 6 distinct analytical categories (e.g., extrema, threshold events, temporal summaries), we synthesize a large-scale evaluation dataset of 7,520 natural language/SQL-pairs. This workload systematically scales along temporal dimensions, spatial granularities, and diverse meteorological variables.
\end{enumerate}

For DAIL-SQL ~\cite{gao2024text} and OpenSearch-SQL~\cite{xie2025opensearch}, two open-source, state-of-the-art Text-to-SQL systems, our framework improves the mean execution accuracy from 12.0\% to 34.6\%, i.e., by a factor of 2.9, and reduces the mean query execution time from 101.7 to 44.4 seconds, i.e., by a factor of 2.3.

\section{Meteorological Data Science}

Meteorological data science relies heavily on large-scale, multidimensional datasets to understand climate patterns, assess environmental impacts, and inform policy. Typical users range from climate scientists and environmental researchers to urban planners and policymakers. Their analytical needs frequently involve complex spatio-temporal queries, such as tracking long-term temperature anomalies across specific administrative regions, correlating localized weather phenomena with land-cover types, or aggregating decadal extreme event frequencies. Enabling declarative, natural language access to this data significantly accelerates the time-to-insight for domain experts and data scientists \cite{braschler2019applied}.

\subsection{Meteorological Datasets and Formats}

Earth observation data is natively distributed in specialized scientific formats such as NetCDF-4 (built on HDF5) or GRIB. For example, the ERA5-Land atmospheric reanalysis provides high-resolution, multi-decadal meteorological variables as continuous multi-dimensional arrays. Similarly, the CORINE Land Cover (CLC) inventory maps ecological zones using hierarchical codes and upscaled raster grids. Integrating these inherently array-based and geographically bound formats into a relational PostgreSQL database is a foundational prerequisite for standard SQL querying.

\subsection{Challenges for Natural Language Interfaces}
\label{sub:challenges}

We identify the following challenges regarding natural language interfaces for meteorological data:

\begin{enumerate}[start=1,label={[\bfseries C\arabic*]}, leftmargin = 2em]

    \item \emph{Location Resolution}. When querying meteorological datasets, users naturally frame spatial constraints using symbolic toponyms (e.g., ``in Thessaly'' or ``across the Canton of Bern'') rather than explicit coordinate geometries. Because the underlying relational database relies on precise mathematical geometries or latitude/longitude coordinates, a standard LLM-based Text-to-SQL system must either hallucinate approximate coordinate bounds or omit the spatial filter entirely. Successfully resolving these logical toponyms requires external grounding against geographic knowledge bases.

       \item \emph{The Symbolic-to-Numeric Gap}. LLMs lack the domain-specific meteorological knowledge required to map natural language concepts to encoded database schema elements. Generating correct SQL requires bridging several semantic disconnects:
       
    \begin{itemize}[leftmargin=0em]
        \item \textbf{Implicit Units:} Meteorological variables from ERA5-Land are stored as floating-point scalars in physical units, such as temperatures in Kelvin, daily precipitation in meters, or solar radiation in J$\cdot$m$^{-2}$. Computing a monthly average temperature in Celsius demands explicit unit conversion (e.g., \texttt{SELECT AVG(tmean) - 273.15}), a convention not derivable from schema metadata alone.
        
        \item \textbf{Array Semantics:} Land-cover classifications follow the CORINE scheme~\cite{clc2018_raster}, mapping hierarchical integer codes to natural language labels. Resolving an entity like \emph{forest} requires structurally valid array subscript predicates to decode the grid-cell pixel counts stored in the database.
        
        \item \textbf{(Composite) Meteorological Concepts:} A query such as \emph{How many heat waves occurred in Athens in 2020} presupposes a formal event definition absent from the schema. Because definitions vary based on local climatology and intensity thresholds~\cite{galanaki2022spatio}, the system must programmatically map abstract concepts to explicit, threshold-duration filtering logic (e.g., $T_{max} \ge 35^\circ\text{C}$ for $\ge 3$ consecutive days).
        
    \end{itemize}

    \item \emph{Computational Complexity of Spatial Queries}. Even when an LLM synthesizes a semantically and structurally correct SQL query, naive execution on several hundred gigabytes of meteorological datasets is prohibitively expensive. Translating broad administrative regions into extensive sets of coordinate pairs often yields massively inflated \texttt{IN}-list predicates. Without query-rewriting strategies—such as dynamically transforming these \texttt{IN}-lists into \texttt{VALUES}-based subqueries or leveraging spatial expression indices—database planning and execution times degrade from seconds to hours, rendering interactive exploration infeasible.

    \item \emph{Grid resolution versus vector GIS}. Spatial engines such as ArcGIS \cite{booth2001getting}, PostGIS \cite{obe2011postgis} and Oracle Spatial \cite{kothuri2004pro} are designed for geodesic buffers, topological predicates, and on-the-fly reprojection --- e.g., \textit{"Which sensors lie within 500 meters of a river?"}, queries that are cumbersome to formulate and inefficient to answer. These extensions introduce native geometry and geography types, spatial indexing structure, and a rich set of operators for topological relationships and distance computation. That is not the query class of this paper, and it is not where the Text-to-SQL failures in [\textbf{C1}]--[\textbf{C3}] occur.

    ERA5-Land~\cite{munoz2021era5} stores each variable on a WGS84 grid with $0.1^\circ$ spacing (roughly $10$\,km, ${\sim}100$\,km$^2$ per cell),  CORINE~\cite{clc2018_raster}, originally at $100$\,m in ETRS89. Before either product is queried, we resample the land-cover raster onto the ERA5-Land lattice and store both as relational fact tables of the form $(\mathrm{time},\,\mathrm{lat},\,\mathrm{lon},\, \mathrm{value})$ plus an upscaled land-cover array. Coordinate systems are therefore reconciled once, when the database is built. 
    
    Figure~\ref{fig:postgis_zh_example} shows that each grid cell holds only one averaged temperature, and no amount of interpolating in PostGIS can split that average back into the hill, the lake and the city it came from. For instance, the cell on the Uetliberg ridge (south west the shaded area of the city of Zurich, at coordinates (8.5, 47.3)) reports $31.1^\circ$C, and the cell on the eastern shore of Lake Zurich (8.6, 47.3) reports $31.0^\circ$C, despite several hundred meters of elevation difference and opposite land cover. Each number is already an average over urban fabric, forest, water or suburbs around it. Interpolating the grid in order to ask whether ``the city is warmer than the surrounding countryside'' would manufacture unreal precision that the observations do not contain.

    Consequently we separate three concerns that GIS textbooks often merge. (i)~\emph{Curvature and geodesic distance} matter for within-distance (buffer) queries---for example, ``which stations lie within $500$\,m of the Limmat river?''. For deciding which ERA5-Land cells ($0.1^\circ$, roughly $9$\,km) belong to a canton, however, they shift the boundary by only a few meters, far below the cell resolution, so they almost never change which cell centers fall inside. (ii)~\emph{Coordinate-system alignment} between ERA5-Land and CORINE is required, and is performed once when the tables are loaded, not inside every generated query. (iii)~\emph{Region membership} --- which cells belong to what place, e.g., canton of Zurich --- is required in almost every question we study. We materialize that membership as a discrete set of ERA5 cell centers that fall inside the OSM multipolygon (exact) or inside its bounding box (approximate). Both variants are first-class gold SQL in Section~\ref{sec:stratos_workload}. 
    
    As an example, let us revisit Figure~\ref{fig:postgis_zh_example}. We can observe that 20 temperature values lie within the polygon of the canton of Zurich (marked in blue), an additional 10 temperature values are found inside the bounding box, but not the polygon (marked in orange). A query using the bounding box therefore draws one third of its values from outside the shape of the canton.
    
    Section~\ref{sec:experiments} shows that switching from the set of polygon points to the bounding box \emph{does} change answers for a large fraction of queries; spatial filtering is therefore not optional. We decline to push a continuous \texttt{ST\_Intersects} (or an equivalent geometry predicate) in the generated SQL, for three reasons. First, the predicate still selects the same $0.1^\circ$ ERA5-Land cells. Intersecting that cell with the multipolygon does not recover the measurement across the mix of ridge, lake and urban fabric that Figure~\ref{fig:postgis_zh_example} shows has already been averaged away. Second, a fractional overlay would not give a better gold answer. Splitting a border cell by area would re-weight the same stored value; it cannot invent a separate temperature for the part of the cell that lies inside the place and the part that does not. The exact gold SQL therefore keeps a cell when its center lies inside the multipolygon, and otherwise drops it. Third, a polygon predicate does not remove the bottleneck in [\textbf{C3}]. A canton or a region still corresponds to tens or hundreds of cells. Written as \texttt{ST\_Intersects} or as a long \texttt{IN}-list of coordinates, the query still has to touch that many grid points. The planner cost we fix with approximate or query re-writer in Section~\ref{sec:experiments} comes from that list cardinality, not from how the places were turned into those pairs.

    Figure~\ref{fig:postgis_zh_example} illustrates the two ways we define a region in SQL. The \textit{exact query} keeps only the cells whose centers fall inside the region's polygon. The \textit{approximate query} keeps every cell whose center falls inside the region's bounding box, and also includes cells that lie outside the region itself. We are aware that the two query variants can return different sets of cells and therefore different answers. Rather than choosing one, we include both as gold SQL and report in Section~\ref{sec:experiments} how often and by how much they disagree. In either case, the filter only decides which cells belong to the region; it does not refine the values within a cell, which remain averages over roughly $10$\,km $\times$ $10$\,km.

    \begin{figure}[htb]
        \centering
        \includegraphics[width=\linewidth]{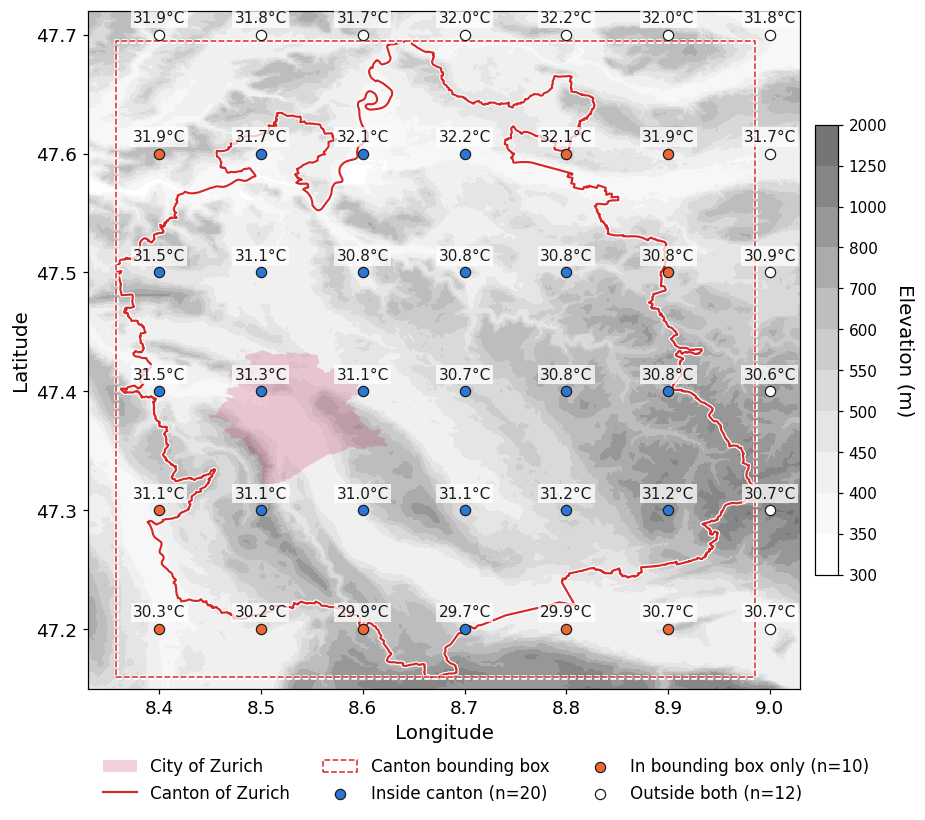}
    \caption{Example of available temperature data points in ERA5-Land~\cite{munoz2021era5}, superimposed over the area of the canton of Zurich, Switzerland. The city of Zurich is marked as a shaded area in the left half of the figure. The grayscale shading corresponds to the elevation. Each dot corresponds to a temperature value, averaged over a cell of 10 km $\times$ 10 km of mixed terrain. The color of each dot indicates whether it lies inside the polygon of the canton of Zurich (blue), only inside the canton's bounding box (orange), or outside both (white).}    \label{fig:postgis_zh_example}
    \rmspace
    \end{figure}

\end{enumerate}

\section{STRATOS: Spatio-Temporal Resolution Agent for Text-to-SQL}

\begin{figure*}[h!]
    \centering
    \includegraphics[width=1.0\linewidth]{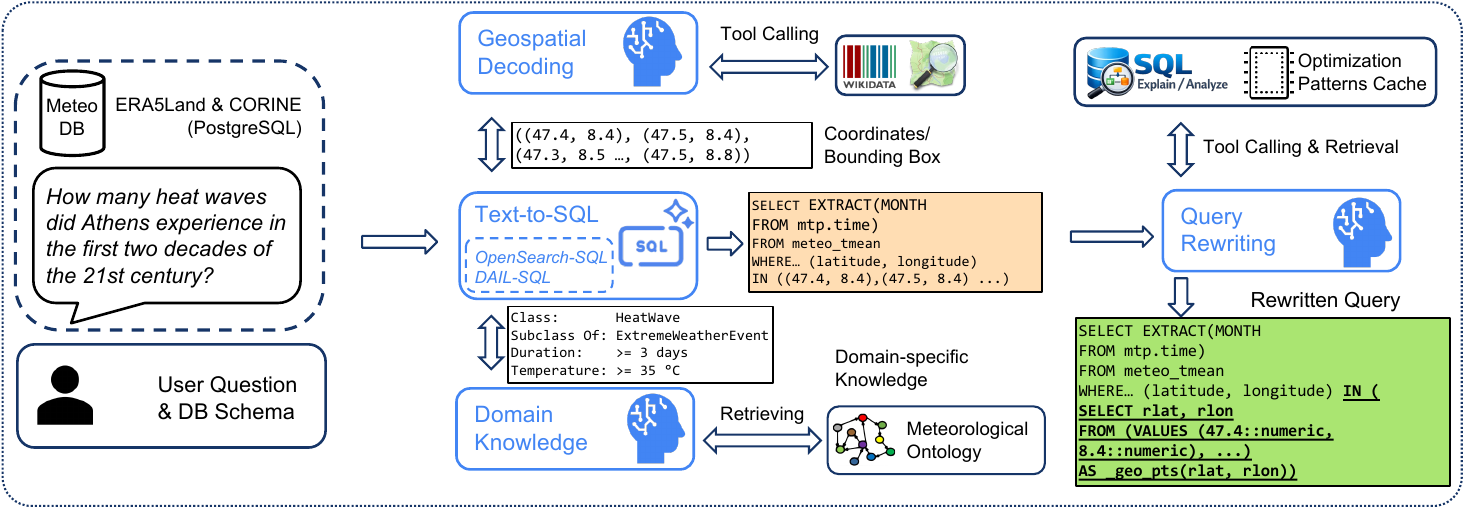}
\caption{Overview of the STRATOS multi-agent architecture. External APIs and domain ontologies augment the generative Text-to-SQL pipeline, followed by automated query rewriting.}    \label{fig:stratos_arch}
\rmspace
\end{figure*}

Figure~\ref{fig:stratos_arch} illustrates the STRATOS architecture, designed to address the specific data science challenges outlined in Section~\ref{sub:challenges}. A user's natural language question is first processed and analyzed by the \textit{Text-to-SQL} component that in turn interacts with the \textit{Geo Decoding} component and the \textit{Domain Knowledge} component. The \textit{Geo Decoding} component resolves symbolic toponyms to spatial coordinates using Wikidata and OpenStreetMap [\textbf{C1}], while a \textit{Domain Knowledge} component retrieves relevant meteorological definitions from an ontology. Using this enriched context, the generative \textit{Text-to-SQL} component synthesizes an initial candidate query [\textbf{C2}]. Finally, a \textit{Query Rewriter} leverages tool-calling and a query cache to refine the candidate query for efficient database execution [\textbf{C3}].

\subsection{Geo-Spatial Context Resolution}

To address spatial ambiguity [\textbf{C1}], we implemented a geospatial retriever that extracts location entities from natural language questions and resolves them into precise geographic geometries using external knowledge bases.
Specifically, the retriever first queries the Wikidata~\cite{vrand2014wiki} knowledge base using the extracted place name. To resolve potential geographical ambiguities, an LLM evaluates the top-$k$ candidate entities returned by Wikidata and selects the most contextually appropriate match. The selected Wikidata Entity Identifier (QID, e.g., \texttt{Q72} for the City of Zurich) is subsequently used to query the OpenStreetMap (OSM) API. This allows the system to retrieve the corresponding GeoJSON data, capturing both the spatial bounding box (minimum and maximum latitude and longitude) and the precise multi-polygon boundary of the target entity. This geometry is then injected into the SQL generation prompt, ensuring accurate spatial filtering downstream.


\subsection{Semantic Translation and Bridging the Symbolic-Numeric Gap}

To bridge the symbolic-to-numeric gap [\textbf{C2}], the Domain Knowledge component employs two complementary mechanisms: ontology grounding for meteorological concepts and array-semantic hints for complex land-cover schemas.

\subsubsection{Ontology Grounding}

We encode formal meteorological logic in an OWL/RDF ontology (\texttt{.ttl}), ensuring definitions remain customizable and distinct from pipeline code (see schematic representation in Figure~\ref{fig:stratos_arch}). During execution, an agent performs a \emph{semantic root search}: it matches the user's question to a top-ranked ontology anchor via embedding similarity, then expands the graph along a controlled predicate list (e.g., \textit{exceedsThreshold}, \textit{hasConversionRule}) to reconstruct the full definition~\cite{bhalotia2002banks,rony2022treekgqa}.
Starting from an anchor node like \textsc{HeatWave}, the system traverses the knowledge graph to extract formal constraints.
The extracted \textsc{HeatWave} subgraph is serialized into an \emph{Ontology Grounding Function}
and appended to the LLM prompt. This mechanism seamlessly handles composite climate indices (e.g., \textsc{FrostDay}, \textsc{TropicalNight}) without additional prompt engineering.

\subsubsection{Array-Semantic Hint}

The CORINE land-cover schema introduces a unique structural hurdle: regional classifications are stored as 2D integer arrays (\texttt{integer[][]}). Translating natural language against this schema requires overcoming two sub-challenges:

    \textbf{Entity Matching:} Natural language terms such as "city" or "lake" must map exactly to the corresponding CORINE labels, namely "urban fabric" or "inland waters", respectively. The agent first applies substring validations (\texttt{ILIKE}) against CORINE labels. For terms lacking exact matches, an embedding-based fallback resolves the discrepancy (e.g., mapping ``lake'' to the broader category ``Inland waters'' ).
    
   \textbf{Semantic Query Structuring:} LLMs struggle to spontaneously generate syntax for multidimensional array parsing. To guide it, a semantic hint module classifies the user's question into one of six pre-defined access paradigms (e.g., \textit{membership filter}, \textit{aggregate distribution}). 
    E.g., querying \textit{``urban-dominated areas''} triggers the \textit{membership filter} paradigm.

\subsection{Query Rewriting} \label{sub:query_rewriting}

The temporal and spatial complexity of meteorological data introduces severe query performance bottlenecks [\textbf{C3}], which we illustrate using the query \emph{``What is the average temperature in Celsius in the Canton of Bern in 2016?''} Geo-spatial resolution maps the Canton of Bern boundary to 65 grid-point pairs. A standard LLM typically generates the following predicate structure:

\begin{Verbatim}[commandchars=\\\{\},fontsize=\footnotesize,breaklines,breakanywhere]
\PYG{k}{SELECT}\PYG{+w}{ }\PYG{k}{AVG}\PYG{p}{(}\PYG{n}{tmean}\PYG{p}{)}\PYG{+w}{ }\PYG{o}{\PYGZhy{}}\PYG{+w}{ }\PYG{l+m+mi}{273}\PYG{p}{.}\PYG{l+m+mi}{15}
\PYG{k}{FROM}\PYG{+w}{ }\PYG{n}{meteo\PYGZus{}tmean}
\PYG{k}{WHERE}\PYG{+w}{ }\PYG{k}{EXTRACT}\PYG{p}{(}\PYG{k}{YEAR}\PYG{+w}{ }\PYG{k}{FROM}\PYG{+w}{ }\PYG{k}{time}\PYG{p}{)}\PYG{+w}{ }\PYG{o}{=}\PYG{+w}{ }\PYG{l+m+mi}{2016}
\PYG{+w}{  }\PYG{k}{AND}\PYG{+w}{ }\PYG{p}{(}\PYG{n}{ROUND}\PYG{p}{(}\PYG{k}{CAST}\PYG{p}{(}\PYG{n}{latitude}\PYG{+w}{  }\PYG{k}{AS}\PYG{+w}{ }\PYG{n+nb}{DECIMAL}\PYG{p}{),}\PYG{+w}{ }\PYG{l+m+mi}{1}\PYG{p}{),}
\PYG{+w}{       }\PYG{n}{ROUND}\PYG{p}{(}\PYG{k}{CAST}\PYG{p}{(}\PYG{n}{longitude}\PYG{+w}{ }\PYG{k}{AS}\PYG{+w}{ }\PYG{n+nb}{DECIMAL}\PYG{p}{),}\PYG{+w}{ }\PYG{l+m+mi}{1}\PYG{p}{))}
\PYG{+w}{  }\PYG{k}{IN}\PYG{+w}{ }\PYG{p}{((}\PYG{l+m+mi}{47}\PYG{p}{.}\PYG{l+m+mi}{1}\PYG{p}{,}\PYG{+w}{ }\PYG{l+m+mi}{6}\PYG{p}{.}\PYG{l+m+mi}{9}\PYG{p}{),}\PYG{+w}{ }\PYG{p}{(}\PYG{l+m+mi}{47}\PYG{p}{.}\PYG{l+m+mi}{0}\PYG{p}{,}\PYG{+w}{ }\PYG{l+m+mi}{7}\PYG{p}{.}\PYG{l+m+mi}{1}\PYG{p}{),}\PYG{+w}{ }\PYG{p}{...)}\PYG{+w}{ }\PYG{c+c1}{\PYGZhy{}\PYGZhy{} 65 pairs}
\end{Verbatim}

\noindent \textbf{Temporal Bottleneck.}
The \texttt{EXTRACT} predicate on a raw timestamp column is opaque to a standard B-tree index, forcing a full sequential scan (with an estimated cost of $\approx 4{,}000{,}000$ according to the PostgreSQL query planner). This is resolved by creating a \textit{multi-column expression index} over the rounded coordinates alongside the extracted year and month:

\begin{Verbatim}[commandchars=\\\{\},fontsize=\footnotesize,breaklines,breakanywhere]
\PYG{k}{CREATE}\PYG{+w}{ }\PYG{k}{INDEX}\PYG{+w}{ }\PYG{k}{ON}\PYG{+w}{ }\PYG{n}{meteo\PYGZus{}tmean}\PYG{+w}{ }\PYG{p}{(}
\PYG{+w}{  }\PYG{n}{ROUND}\PYG{p}{(}\PYG{n}{latitude}\PYG{p}{::}\PYG{n+nb}{numeric}\PYG{p}{,}\PYG{+w}{ }\PYG{l+m+mi}{1}\PYG{p}{),}\PYG{+w}{ }\PYG{n}{ROUND}\PYG{p}{(}\PYG{n}{longitude}\PYG{p}{::}\PYG{n+nb}{numeric}\PYG{p}{,}\PYG{+w}{ }\PYG{l+m+mi}{1}\PYG{p}{),}
\PYG{+w}{  }\PYG{k}{EXTRACT}\PYG{p}{(}\PYG{k}{YEAR}\PYG{+w}{ }\PYG{k}{FROM}\PYG{+w}{ }\PYG{k}{time}\PYG{p}{),}\PYG{+w}{     }\PYG{k}{EXTRACT}\PYG{p}{(}\PYG{k}{MONTH}\PYG{+w}{ }\PYG{k}{FROM}\PYG{+w}{ }\PYG{k}{time}\PYG{p}{));}
\end{Verbatim}

This index allows the planner to satisfy the \texttt{EXTRACT} condition directly as an index predicate, reducing the cost by a factor of 100. This expression index is proactively applied to all meteorological fact tables in our database.

\noindent \textbf{Spatial Bottleneck.}
Even with the expression index, the 65-pair \texttt{IN}-list prevents a single index lookup. The query planner decomposes it into 65 independent \texttt{Bitmap Index Scan} nodes merged via \texttt{BitmapOr}, paired with a \texttt{Recheck} condition growing linearly with the number of coordinate pairs. 
To resolve this, the Query Rewriter rewrites the \texttt{IN}-list into a \texttt{VALUES}-based subquery:

\begin{Verbatim}[commandchars=\\\{\},fontsize=\footnotesize,breaklines,breakanywhere]
\PYG{+w}{  }\PYG{k}{AND}\PYG{+w}{ }\PYG{p}{(}\PYG{n}{ROUND}\PYG{p}{(}\PYG{k}{CAST}\PYG{p}{(}\PYG{n}{latitude}\PYG{+w}{  }\PYG{k}{AS}\PYG{+w}{ }\PYG{n+nb}{DECIMAL}\PYG{p}{),}\PYG{+w}{ }\PYG{l+m+mi}{1}\PYG{p}{),}
\PYG{+w}{       }\PYG{n}{ROUND}\PYG{p}{(}\PYG{k}{CAST}\PYG{p}{(}\PYG{n}{longitude}\PYG{+w}{ }\PYG{k}{AS}\PYG{+w}{ }\PYG{n+nb}{DECIMAL}\PYG{p}{),}\PYG{+w}{ }\PYG{l+m+mi}{1}\PYG{p}{))}\PYG{+w}{ }\PYG{k}{IN}\PYG{+w}{ }\PYG{p}{(}
\PYG{+w}{    }\PYG{k}{SELECT}\PYG{+w}{ }\PYG{n}{rlat}\PYG{p}{,}\PYG{+w}{ }\PYG{n}{rlon}
\PYG{+w}{    }\PYG{k}{FROM}\PYG{+w}{ }\PYG{p}{(}\PYG{k}{VALUES}\PYG{+w}{ }\PYG{p}{(}\PYG{l+m+mi}{47}\PYG{p}{.}\PYG{l+m+mi}{1}\PYG{p}{::}\PYG{n+nb}{numeric}\PYG{p}{,}\PYG{+w}{ }\PYG{l+m+mi}{6}\PYG{p}{.}\PYG{l+m+mi}{9}\PYG{p}{::}\PYG{n+nb}{numeric}\PYG{p}{),}\PYG{+w}{ }\PYG{p}{...)}
\PYG{+w}{      }\PYG{k}{AS}\PYG{+w}{ }\PYG{n}{\PYGZus{}geo\PYGZus{}pts}\PYG{p}{(}\PYG{n}{rlat}\PYG{p}{,}\PYG{+w}{ }\PYG{n}{rlon}\PYG{p}{))}
\end{Verbatim}

Despite a relatively modest difference in estimated query cost ($7{,}632$ vs.\ $6{,}569$), the actual execution times on a cold buffer cache reveal a substantial performance gap: \textbf{2{,}287\,ms} for the \texttt{BitmapOr} plan compared to just \textbf{32\,ms} for the rewritten query---\textbf{yielding an end-to-end speedup of} $\mathbf{68\times}$ (Table~\ref{tab:optimizer}). The divergence is explained by the structural difference in the two query execution plans:

\noindent\textbf{Before rewrite} -- \texttt{BitmapOr} within \texttt{Bitmap Heap Scan}:
\begin{Verbatim}[commandchars=\\\{\},fontsize=\footnotesize,bgcolor=Salmon!20,breaklines]
Aggregate  (cost=7632)  (actual time=2284 ms)
  \PYGZhy{}\PYGZgt{} Bitmap Heap Scan on meteo\PYGZus{}tmean
       Recheck Cond: (... 65\PYGZhy{}clause OR condition ...)
       Heap Blocks: exact=2249
       \PYGZhy{}\PYGZgt{} BitmapOr
            \PYGZhy{}\PYGZgt{} Bitmap Index Scan x65 on idx\PYGZus{}tmean\PYGZus{}round\PYGZus{}latlon\PYGZus{}ym
                 (cost=4.86 each)  (actual time=\PYGZti{}0.1 ms each)
...
Planning Time:  19.816 ms
Execution Time: 2287.316 ms
\end{Verbatim}

\noindent\textbf{After rewrite} -- \texttt{Nested Loop} over \texttt{Values Scan}
:
\begin{Verbatim}[commandchars=\\\{\},fontsize=\footnotesize,bgcolor=OliveGreen!15,breaklines]
Aggregate  (cost=6569)  (actual time=23 ms)
  \PYGZhy{}\PYGZgt{} Nested Loop
       \PYGZhy{}\PYGZgt{} HashAggregate on Values Scan
            (65 rows, 24 kB memory)
       \PYGZhy{}\PYGZgt{} Index Scan x65 on idx\PYGZus{}tmean\PYGZus{}round\PYGZus{}latlon\PYGZus{}ym
            (cost=97 each)  (actual time=0.26 ms each)
...
Planning Time:   1.908 ms
Execution Time: 31.867 ms
\end{Verbatim}

\begin{table}[htbp]
\caption{Query execution times for the Canton of Bern temperature query before and after query rewriting.}
\footnotesize
\centering
\begin{tabular}{lrrr}
\toprule
 & \textbf{Planning} & \textbf{Execution} & \textbf{Total} \\
\midrule
\texttt{IN}-list (\texttt{BitmapOr}) & 19.8\,ms & 2{,}287\,ms & 2{,}307\,ms \\
\texttt{VALUES} subquery             &  1.9\,ms &      32\,ms &      34\,ms \\
Speedup                              & $10\times$ & $71\times$ & $\mathbf{68\times}$ \\
\bottomrule
\end{tabular} 
\label{tab:optimizer}
\end{table}

To automate such performance gains, the Query Rewriter detects large coordinate \texttt{IN}-list predicates via regular expression matching and refactors them into \texttt{VALUES}-subqueries. Each rewritten query is then verified for correctness by executing an EXPLAIN statement against the target database. 
Together, the \textbf{temporal expression indexing and spatial query rewriting reduce execution time by up to two orders of magnitude on region-scale queries}. For instance, the original unindexed LLM query times out after 2 hours, whereas the rewritten query executes in 2.3 seconds---representing a speed-up of ${\approx}3{,}200\times$.

\noindent\textbf{DBMS-specifc Query Rewriter}. However, the effectiveness of this query rewriter is DBMS-dependent. In a set of comparative experiments, we test the behaviors of query planners on two additional popular DBMSes, namely \textit{DuckDB} and \textit{SQLite}. 

DuckDB's optimizer produces equivalent logical plans for both (original/rewritten) SQL statements and ultimately selects the same physical plan, i.e., a \texttt{hash semi-join}, so the source-level \texttt{VALUES} wrapper is redundant. SQLite does not apply an analogous rewrite: both the row-constructor \texttt{IN} list and \texttt{IN (SELECT \ldots FROM VALUES)} remain a \texttt{SCAN} of the fact table plus a \texttt{LIST SUBQUERY} over constant rows, and neither form attaches the expression index. A useful optimization on SQLite must expose an explicit join to \texttt{VALUES}, whose plan is a \texttt{SCAN} of $N$ constant rows followed by \texttt{SEARCH} on that index, rather than another \texttt{IN} spelling. 

In summary, both engines therefore show no gain from the query rewriting suggested for PostgreSQL. Moreover, DuckDB has already performed a rewrite of that kind while SQLite does not.

\section{The STRATOS Evaluation Workload}
\label{sec:stratos_workload}

To evaluate our spatio-temporal framework, we created a specialized Text-to-SQL workload designed to stress-test both the symbolic-to-numeric reasoning of LLMs and the execution scalability of the underlying database.

\subsection{Workload Generation and Scaling}

The evaluation workload is built upon a foundation of 161 manually curated, domain-expert-verified natural language questions spanning six distinct analytical categories (e.g., temporal summaries, extrema, anomalies). To rigorously evaluate system scalability across varied computational regimes, we abstracted these seeds into 376 parameterized templates. We systematically sampled parameters from predefined pools---including 7 climate variables, multiple temporal scales (from daily to decadal), and 20 geographically diverse locations across Switzerland and Greece. 

A critical feature of this workload is its \textit{dual-spatial instantiation}, i.e., each natural language query is instantiated twice, once with an exact point-wise spatial filter and once with an approximate bounding-box filter, yielding two gold-standard SQL variants to test different database execution profiles:
\begin{itemize}[leftmargin=1em]
    \item \textbf{Point-wise Queries (Exact):} These enforce precise multi-polygon boundaries by checking intersections against hundreds of explicit coordinate pairs. While highly accurate, they are computationally intensive and frequently trigger out-of-memory exceptions or hours-long execution times in unoptimized systems.
    \item \textbf{Bounding-Box Queries (Approximate):} These simplify the spatial constraint into an encompassing rectangle. The rectangle is easier to express than an explicit list of coordinate pairs, but it covers more grid cells than the exact point set and thus trades off geographical precision (see also example in Figure~\ref{fig:postgis_zh_example}).
\end{itemize}

By including both variants, the benchmark yields a total of 7,520 NL/SQL-pairs (summarized in Table~\ref{tab:category-taxonomy}), enabling a controlled comparison of query rewriting effectiveness across exact and approximate spatial filtering.

\begin{table}[htbp]
  \centering
  \caption{STRATOS Benchmark Queries.}
  \label{tab:category-taxonomy}
  \small
  \begin{tabular}{@{}llr@{}}
    \toprule
    \textbf{Cat.} & \textbf{Description}  & \textbf{NL/SQL-Pairs} \\
    \midrule
    AGG & Aggregates \& Temporal Summaries   & 1{,}580 \\
    CHG & Change, Trend \& Comparison         &   840 \\
    EVT & Events, Thresholds \& Persistence   & 1{,}200 \\
    EXT & Extrema \& Conditional Extrema     & 2{,}440 \\
    ORD & Ranking \& Distribution             &   580 \\
    SPC & Spatial \& Elevation Constraints    &   880 \\
    \midrule
    & \textbf{Total} & \textbf{7{,}520} \\
    \bottomrule
  \end{tabular}
  \rmspace
\end{table}

\subsection{Decoupled Workload Hardness}

Standard Text-to-SQL benchmarks such as Spider~\cite{yu-etal-2018-spider} typically approximate query difficulty by counting the number of constraints (e.g., \texttt{WHERE} clauses). In scalable data science, this heuristic fails. A meteorological query spanning ten years across an entire country may require massive database processing power but possess a relatively simple SQL structure. Conversely, extracting a complex array-based land-cover metric for a single grid cell is semantically difficult for an LLM but executes in fractions of a second. 

To resolve this contradiction, we annotate every query in the workload with \textit{two decoupled hardness scores} (see Table~\ref{tab:dataset-splits-transposed}). Both scores are computed from the gold SQL alone, which we parse with \texttt{SQLGlot}, and are therefore agnostic to the natural language question and the data rows.

\noindent\textbf{Semantic Hardness} measures the structural complexity of schema linking:
\begin{equation}
    \label{eq:semantic-score}
    \mathcal{H}_{\mathrm{sem}} = \omega_{m} T_{m} + \omega_{x} T_{x} + \omega_{j} J + \omega_{n} N + \omega_{a} A ,
\end{equation}
where $T_{m}$ and $T_{x}$ count the distinct meteorological tables (e.g., mean temperature, total precipitation) and auxiliary tables (e.g., elevation, land cover), $J$ counts explicit and implicit joins, $N$ counts CTEs and nested subqueries, and $A \in \{0,1\}$ flags array operations such as \texttt{UNNEST}. Following the component-based hardness criteria of Spider~\cite{yu-etal-2018-spider}, we weigh each component by its expected difficulty for schema linking and set $\omega_{m}{=}2$, $\omega_{x}{=}3$, $\omega_{j}{=}4$, $\omega_{n}{=}2$ and $\omega_{a}{=}8$. 

Auxiliary tables are often more implicit to link than meteorological ones; joins require the most precise alignment of predicates across tables, and array operations are rare and syntactically brittle in standard SQL. Spatial and temporal span are deliberately left out, since a longer period or a larger region only changes the values in the filter predicates, not the complexity of the filter structure. 

A query is \texttt{easy} if $\mathcal{H}_{\mathrm{sem}} \le \tau_{1}^{\mathrm{sem}}$, \texttt{medium} if $\tau_{1}^{\mathrm{sem}} < \mathcal{H}_{\mathrm{sem}} \le \tau_{2}^{\mathrm{sem}}$, and \texttt{hard} otherwise, with $\tau_{1}^{\mathrm{sem}}{=}4$ and $\tau_{2}^{\mathrm{sem}}{=}10$. 
Like Spider~\cite{yu-etal-2018-spider}, these levels are derived from the SQL components of the gold query. We extend Spider's components by CTEs, array operations, and the type of the accessed tables, and weight the components instead of only counting them.

\noindent\textbf{Execution Hardness} estimates the volume of data a query has to process. For exact queries ($v{=}\mathrm{exact}$) and approximate queries ($v{=}\mathrm{approx}$) we compute
\begin{equation}
    \label{eq:exec-score}
    \mathcal{H}_{\mathrm{exec}}^{v} = W_t + \omega_{j} J + \omega_{n} N + \log_{2}(1 + U_{v}),
\end{equation}
where $W_t$ encodes the temporal grain of the query, and $U_{v}$ is the number of spatial units the query reads, i.e., $U_{\mathrm{exact}}$ counts the distinct grid points listed in the query and $U_{\mathrm{approx}}$ the $0.1^{\circ}$ grid cells inside its bounding box. We set $W_t = 1, 2, 3, 4, 5$ for a day, a week, a month, a year, and a cross-year period. The logarithm prevents large regions from dominating the score. The join and nesting terms in Eq.~\ref{eq:semantic-score} use the same weights because each join and nesting level adds more complexity to the querying. 

A question is assigned the higher of its two scores, $\mathcal{H}_{\mathrm{exec}} = \max(\mathcal{H}_{\mathrm{exec}}^{\mathrm{exact}}, \mathcal{H}_{\mathrm{exec}}^{\mathrm{approx}})$. Hence,  its exact and approximate queries share one level: \texttt{small} if $\mathcal{H}_{\mathrm{exec}} \le \tau_{1}^{\mathrm{exec}}$, \texttt{medium} if $\tau_{1}^{\mathrm{exec}} < \mathcal{H}_{\mathrm{exec}} \le \tau_{2}^{\mathrm{exec}}$, and \texttt{large} otherwise, with $\tau_{1}^{\mathrm{exec}}{=}5$ and $\tau_{2}^{\mathrm{exec}}{=}10$. For example, a yearly query on a single grid point scores $5$ and is \texttt{small}, whereas a multi-year aggregate over a region of about $100$ grid cells scores above $11$ and is \texttt{large}.

This decoupled annotation allows us to independently evaluate a system's ability to navigate complex schemas ($\rightarrow$Semantic Hardness) versus its ability to rewrite and optimize large-scale aggregations ($\rightarrow$Execution Hardness).


\begin{table}[htbp]
  \centering
  \caption{Data split: Distribution across Semantic and Execution Hardness levels, \#Places and NL/SQL-pair counts.}
  \label{tab:dataset-splits-transposed}
  \small
    \begin{tabular}{@{}llccc@{}}
    \toprule
     &  & \textbf{Train} & \textbf{Test} & \textbf{Total} \\
    \midrule
    \multirow{3}{*}{Semantic} & Easy & 3{,}744 {\scriptsize (62.2\%)} & 1{,}016 {\scriptsize (67.6\%)} & \textbf{4{,}760} {\scriptsize (63.3\%)} \\
     & Medium & 928 {\scriptsize (15.4\%)} & 224 {\scriptsize (14.9\%)} & \textbf{1{,}152} {\scriptsize (15.3\%)} \\
     & Hard & 1{,}344 {\scriptsize (22.3\%)} & 264 {\scriptsize (17.6\%)} & \textbf{1{,}608} {\scriptsize (21.4\%)} \\
    \midrule
    \multirow{3}{*}{Execution} & Small & 1{,}210 {\scriptsize (20.1\%)} & 70 {\scriptsize (4.7\%)} & \textbf{1{,}280} {\scriptsize (17.0\%)} \\
     & Medium & 1{,}978 {\scriptsize (32.9\%)} & 434 {\scriptsize (28.9\%)} & \textbf{2{,}412} {\scriptsize (32.1\%)} \\
     & Large & 2{,}828 {\scriptsize (47.0\%)} & 1{,}000 {\scriptsize (66.5\%)} & \textbf{3{,}828} {\scriptsize (50.9\%)} \\
     \midrule
    \multicolumn{2}{l}{\#Places} & 16 & 4 & \textbf{20} \\
    \multicolumn{2}{l}{Total Pairs} & 6{,}016 & 1{,}504 & \textbf{7{,}520} \\
    \bottomrule
  \end{tabular}
\end{table}

\subsection{Template-based Data Splitting}


The test set ($1{,}504$ queries) is heavily isolated (see Table~\ref{tab:dataset-splits-transposed}): it comprises entirely held-out examples evaluated against four strictly unseen geographic regions (Zurich, Attica, Canton of Bern, and Peloponnese). The training set ($6{,}016$ queries) utilizes the remaining templates and regions. We carefully balance the sampling distribution to ensure that the ratio of Semantic Hardness (easy/medium/hard) remains consistent across both the training and testing sets, ensuring a fair evaluation baseline.

\section{Experimental Evaluation}
\label{sec:experiments}

Our experimental evaluation is designed to answer three core research questions (RQs) that assess both the natural language comprehension and the database scalability of our framework:

\begin{itemize}[leftmargin=1em]
    \item \textbf{RQ1 (Semantic Translation):} How effectively does STRATOS bridge the symbolic-to-numeric gap compared to state-of-the-art prompt-based and agentic Text-to-SQL baselines?
    \item \textbf{RQ2 (Execution Scalability):} How does the query rewriting impact execution latency, and how do exact (point-wise) queries compare against approximate (bounding box) queries at scale?
    \item \textbf{RQ3 (Ablation):} What is the individual contribution of the geospatial, ontological, and array-semantic agents to the overall system accuracy and scalability?
\end{itemize}

\subsection{Experimental Setup}

\textbf{Data \& Infrastructure.} The meteorological datasets (ERA5-Land and CORINE Land Cover) are hosted on a PostgreSQL v17.7 database. All query execution experiments were conducted on a dedicated server equipped with 16 CPU cores, 32 GB of RAM, and 2000 GB of NVMe storage to ensure consistent buffer cache profiling and execution latency measurements. The database occupies 587\,GB; each of its eight meteorological tables holds about 183 million rows.

\noindent\textbf{Baselines \& Integration Frameworks.} To demonstrate the generalizability and impact of our architecture, we evaluate STRATOS as a capability augmentation over two representative, state-of-the-art, open-source Text-to-SQL systems, each integrated across two distinct, open-source Large Language Model (LLM) backbones---namely, the \textit{Llama-3.3-70B} and \textit{Qwen3-Coder-Next-80B} models:
    (1) \textit{DAIL-SQL~\cite{gao2024text}:} A leading prompt-based Text-to-SQL framework that utilizes advanced few-shot in-context learning and schema-linking heuristics.
    (2) \textit{OpenSearch-SQL~\cite{xie2025opensearch}:} A recent, highly-capable agentic Text-to-SQL system designed for complex reasoning, tool-calling, and iterative query generation.

For each framework and backbone LLM combination, we establish a ``Vanilla'' baseline (the system operating purely on the database DDL and standard schema information) and compare it against STRATOS. This paired evaluation strictly isolates the performance gains directly attributable to our Spatial Resolution, Domain Knowledge Injection, and Query Rewriter under different model capacities and architectures.

\noindent\textbf{Metrics.} We evaluate over the STRATOS Evaluation Workload (Section~\ref{sec:stratos_workload}) using the following primary metrics:
\begin{itemize}[leftmargin=1em]
    \item \textit{Execution Accuracy (EX):} The percentage of generated queries whose execution result set exactly matches the gold-standard.
    \item \textit{Valid Execution Score (VES):} A BIRD benchmark metric~\cite{li2024can} combining execution accuracy and relative query efficiency, penalizing correct but computationally inefficient queries.
    \item \textit{Inference Time:} Time required to process the NL question and generate the final SQL query (in seconds).
    \item \textit{Successful DB Runtime:} The database execution time of the generated SQL queries that successfully completed (in seconds).
    \item \textit{Timeout/OOM Rate:} The percentage of structurally correct queries that failed to execute due to Out-Of-Memory (OOM) exceptions or exceeded a hard timeout limit of 600 seconds (10 minutes).
\end{itemize}

\subsection{RQ1: End-to-End Accuracy and Efficiency}

To evaluate how effectively our system bridges the symbolic-to-numeric gap, we analyze both Execution Accuracy (EX) and Valid Execution Score (VES) across the three semantic hardness levels. 

As shown in Table~\ref{tab:execution_accuracy}, vanilla LLM baselines struggle severely, achieving an execution accuracy (EX) of less than 20\% on medium and less than 10\% on hard queries, respectively. This failure stems directly from their inability to spontaneously synthesize meteorological unit conversions (e.g., Kelvin to Celsius), decipher derived ontology concepts (e.g., ``heat waves''), or generate valid multidimensional array unnesting logic for land-cover data.

Conversely, integrating STRATOS maintains a high execution accuracy across all semantic complexities for both DAIL-SQL and OpenSearch-SQL, across both the Llama and Qwen backbone models. We can observe the highest overall EX improvement of STRATOS from 10.64\% to 51.33\% for OpenSearch-SQL using Qwen3-Coder. \textbf{This corresponds to an EX improvement of almost a factor of 5 over the Vanilla implementation}. 

Furthermore, the high Valid Execution Scores (VES) indicate that STRATOS does not just generate correct queries, but it generates \textit{efficient} queries that align well with gold-standard performance. For instance, the highest overall VES improvement from 0.087 to 0.501 is again achieved with OpenSearch-SQL using Qwen3-Coder. By dynamically injecting Ontology Grounding Functions (OGF) and array-semantic hints into the prompt context, the system successfully shields the generative model from the underlying schema's structural idiosyncrasies. 

\begin{table}[htbp]
  \centering
    \caption{Execution Accuracy (EX) and Valid Execution Score (VES) across semantic hardness levels evaluated under an embedding-similarity-based 3-shot setting. In total 1,504 exact \& approx. queries.}  \label{tab:execution_accuracy}
  \resizebox{\columnwidth}{!}{%
  \begin{tabular}{@{}llcccccc|cc@{}}
    \toprule
    \multirow{2}{*}{\textbf{LLM-Backbone}} & \multirow{2}{*}{\textbf{System Config.}} & \multicolumn{2}{c}{\textbf{Easy}} & \multicolumn{2}{c}{\textbf{Medium}} & \multicolumn{2}{c}{\textbf{Hard}} & \multicolumn{2}{c}{\textbf{Overall}} \\
    & & EX & VES & EX & VES & EX & VES & EX & VES \\
    \midrule
    \multirow{4}{*}{\textbf{\shortstack{Llama-3.3-\\70B}}} & DAIL-SQL (Vanilla) & 17.03\% & 0.126 & 13.39\% & 0.103 & 6.06\% & 0.058 & 14.56\% & 0.111 \\
     & \textbf{STRATOS$_{DAIL-SQL}$} & \textbf{26.18\%} & \textbf{0.233} & \textbf{25.00\%} & \textbf{0.245} & \textbf{8.33\%} & \textbf{0.074} & \textbf{22.87\%} & \textbf{0.207} \\
    \cmidrule(l){2-10}
     & OpenSearch-SQL (Vanilla) & 9.65\% & 0.085 & 9.38\% & 0.084 & 0.38\% & 0.004 & 7.98\% & 0.071 \\
     & \textbf{STRATOS$_{OpenSearch-SQL}$} & \textbf{19.19\%} & \textbf{0.183} & \textbf{12.05\%} & \textbf{0.119} & \textbf{0.76\%} & \textbf{0.008} & \textbf{14.89\%} & \textbf{0.142} \\
    \midrule
    \multirow{4}{*}{\textbf{\shortstack{Qwen3-Coder-\\Next-80B}}} & DAIL-SQL (Vanilla) & 16.54\% & 0.126 & 16.07\% & 0.122 & 7.20\% & 0.067 & 14.83\% & 0.115 \\
     & \textbf{STRATOS$_{DAIL-SQL}$} & \textbf{57.19\%} & \textbf{0.513} & \textbf{51.34\%} & \textbf{0.493} & \textbf{18.18\%} & \textbf{0.169} & \textbf{49.47\%} & \textbf{0.450} \\
    \cmidrule(l){2-10}
     & OpenSearch-SQL (Vanilla) & 12.01\% & 0.095 & 13.39\% & 0.119 & 3.03\% & 0.030 & 10.64\% & 0.087 \\
     & \textbf{STRATOS$_{OpenSearch-SQL}$} & \textbf{63.88\%} & \textbf{0.621} & \textbf{36.16\%} & \textbf{0.359} & \textbf{15.91\%} & \textbf{0.159} & \textbf{51.33\%} & \textbf{0.501} \\
    \bottomrule
  \end{tabular}%
  }
    \rmspace
\end{table}

\subsection{RQ2: Execution Scalability and Rewriter Impact}

Generating a semantically correct SQL query is only half the challenge in meteorological data science. While the inference time (the LLM generation phase) remains relatively constant regardless of query scale, executing the generated spatial queries against large raster arrays dictates the true DB runtime bottleneck.

\noindent\textbf{Query Rewriting Impact.} The STRATOS Query Rewriter bridges this DB runtime performance gap. As detailed in Section~\ref{sub:query_rewriting}, unoptimized point-wise queries rely on massive coordinate \texttt{IN}-lists that force the query planner to generate highly inefficient \texttt{BitmapOr} scans. By automatically rewriting these spatial filters into \texttt{VALUES}-based subqueries and leveraging multi-column expression indices, STRATOS achieves dramatic speedups. 

For selected queries with one-shot, the rewriter reduces the DB runtime from 9.763 seconds to 0.199 seconds - an improvement of $49.2\times$. Furthermore, in the 3-shot setting, the rewriter reduces the system's overall timeout rate from 5.6\% to 4.4\%, demonstrating that STRATOS is highly viable for interactive data exploration. 

\noindent\textbf{Exact vs. Approximate Queries on Gold SQL.} To evaluate the impact of the query rewriter independently from the LLM inference quality, we directly analyze the gold SQL queries.
We use a 6-hour timeout to isolate and measure the before/after-query-rewriting effect on all exact queries, and compare the 717 questions whose original exact and approximate queries both finish.

\begin{figure}[htbp]
  \centering
  \includegraphics[width=\linewidth]{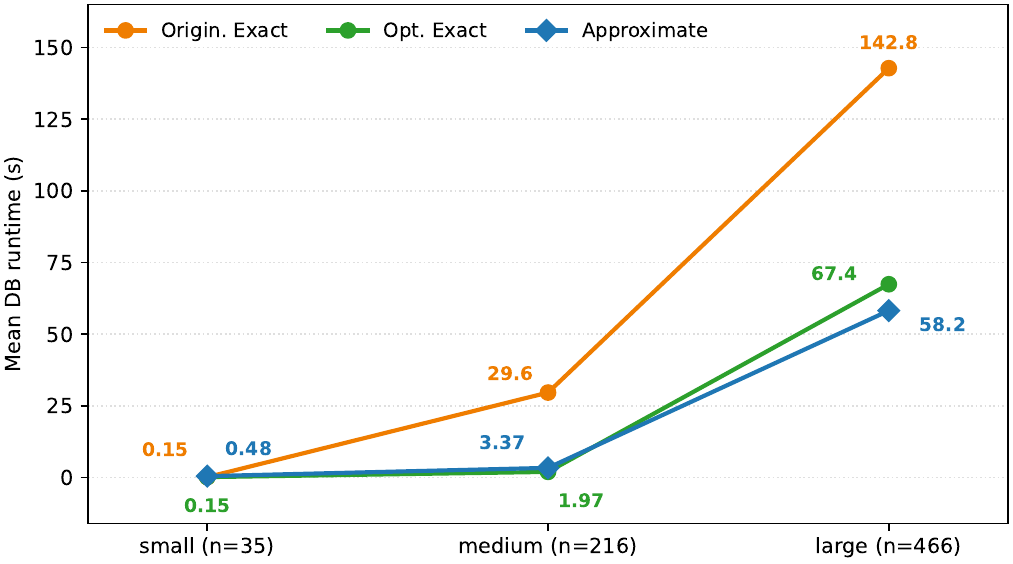}
  \caption{Mean DB runtime of the gold SQL across execution hardness levels for original exact, Optimized exact, and approximate queries.}
  \label{fig:scalability_plot_gold_sql}
    \rmspace
\end{figure}

Figure~\ref{fig:scalability_plot_gold_sql} shows the mean DB runtime for the three execution hardness levels. The mean runtime of the \textit{original exact queries} grows from 0.15 seconds for small queries to 142.8 seconds for large ones. For \textit{optimized exact queries}, the mean runtime for large queries drops to 67.4 second - more than a factor of 2 improvement over the original queries. 

\textit{Approximate queries} are faster than the original exact queries for medium and large queries, with a mean of 3.37 and 58.2 seconds, respectively. For large queries, approximate queries are also faster than optimized queries.

In general, approximate queries are typically faster than exact queries since the bounding box replaces the long list of coordinate pairs (whose cost we discussed in Section~\ref{sub:query_rewriting}) with a single range predicate. 
Although the bounding box covers more grid cells than the exact point set, checking one range is cheaper than matching every row against a long list of coordinate pairs. 

\noindent\textbf{Query Rewriting on Gold SQL.} Let us now analyze the impact of rewriting exact queries. Small queries are not rewritten and keep their runtime. For medium queries, rewriting reduces the mean runtime from 29.6 to 1.97 seconds, i.e., by a factor of 15. For large queries,  query rewriting reduces the run time from 142.8 to 67.4 seconds, i.e., by a factor of more than 2. 

Moreover, 27 of the original exact queries time out, while none of the rewritten queries do (see Table~\ref{tab:optimizer_speedup_gold_sql_pretty}). Over all 717 questions, rewriting reduces the total runtime of the exact queries from 20.3 to 8.8 hours, close to the 7.7 hours of the approximate queries. 


In summary, query rewriting reduces the mean runtime of exact queries by a factor of 15 and 2 for medium and large queries and avoids timeouts (queries that run longer than 6 hours). Approximate queries remain the fastest option for large queries, but at the cost of precision - as we will analyze in the next section (see also Table~\ref{tab:gold_precision_loss_tolerance}). The choice between exact and approximate queries in STRATOS is therefore a trade-off between precision and runtime.

\begin{table}[htbp]
  \centering
  \caption{DB runtime of the gold SQL by execution hardness level: mean and total runtime over the questions whose original exact and approximate queries both finish ($n$), and the number of timeouts over all 752 questions. Opt. = rewritten exact queries.}
  \footnotesize
  \label{tab:optimizer_speedup_gold_sql_pretty}
  \resizebox{\columnwidth}{!}{%
  \begin{tabular}{@{}lr ccc ccc ccc@{}}
    \toprule
    \multirow{2}{*}{\textbf{Lv.}} & \multirow{2}{*}{$\boldsymbol{n}$} &
    \multicolumn{3}{c}{\textbf{Mean (seconds)}} & \multicolumn{3}{c}{\textbf{Total (hours)}} & \multicolumn{3}{c}{\textbf{Timeout (n)}} \\
    \cmidrule(lr){3-5}\cmidrule(lr){6-8}\cmidrule(lr){9-11}
     & & \textbf{Orig.} & \textbf{Opt.} & \textbf{Appr.} & \textbf{Orig.} & \textbf{Opt.} & \textbf{Appr.} & \textbf{Orig.} & \textbf{Opt.} & \textbf{Appr.} \\
     \midrule
    S & 35  & 0.151 & 0.151 & 0.476 & $<$0.01 & $<$0.01 & $<$0.01 & 0  & 0 & 0  \\
    M & 216 & 29.643 & 1.969 & 3.369 & 1.78  & 0.12 & 0.20 & 1  & 0 & 0  \\
    L & 466 & 142.791 & 67.426 & 58.212 & 18.48 & 8.73 & 7.54 & 26 & 0 & 11 \\
    \midrule
    \textbf{All} & \textbf{717} & \textbf{101.741} & \textbf{44.423} & \textbf{38.872} & \textbf{20.26} & \textbf{8.85} & \textbf{7.74} & \textbf{27} & \textbf{0} & \textbf{11} \\
    \bottomrule
  \end{tabular}
  }
\end{table}

\begin{table}[htbp]
  \centering
  \caption{Gold SQL answer agreement between exact (point) and approximate (bounding box) geo filters as a function of the accepted error. At tolerance level $n$ (percent) the per-value budget is $\epsilon = \max(n\% \times |v_{\mathrm{exact}}|,\ 0.01n)$, and two results agree if they share the same shape and every numeric pair satisfies $|v_{\mathrm{exact}} - v_{\mathrm{approx}}| \leq \epsilon$; the absolute floor scales with $n$ so that near-zero quantities (precipitation, snow) remain comparable. Results that differ structurally (row/column shape, or empty approximate result) count as disagreement at every tolerance. \textit{Irreducible Loss} is the residual disagreement at the 10\% practical ceiling.}
  \label{tab:gold_precision_loss_tolerance}
  \scriptsize
  \setlength{\tabcolsep}{4pt}
  \begin{tabular}{@{}llrrrrr@{}}
  \toprule
   & \textbf{Level} & \textbf{N} & \textbf{Agree@1\%} & \textbf{Agree@5\%} & \textbf{Agree@10\%} & \textbf{\shortstack{Irreducible\\Loss}} \\
  \midrule
  \multirow{3}{*}{\shortstack[l]{\tiny Semantic \\ \tiny Hardness}} & Easy & 508 & 247 (48.6\%) & 332 (65.4\%) & 372 (73.2\%) & \textbf{136 (26.8\%)} \\
   & Medium & 112 & 40 (35.7\%) & 56 (50.0\%) & 62 (55.4\%) & \textbf{50 (44.6\%)} \\
   & Hard & 132 & 21 (15.9\%) & 55 (41.7\%) & 66 (50.0\%) & \textbf{66 (50.0\%)} \\
  \midrule
  \multirow{3}{*}{\shortstack[l]{\tiny Execution \\ \tiny Hardness}} & Small & 35 & 19 (54.3\%) & 25 (71.4\%) & 28 (80.0\%) & \textbf{7 (20.0\%)} \\
   & Medium & 217 & 101 (46.5\%) & 147 (67.7\%) & 162 (74.7\%) & \textbf{55 (25.3\%)} \\
   & Large & 500 & 188 (37.6\%) & 271 (54.2\%) & 310 (62.0\%) & \textbf{190 (38.0\%)} \\
  \midrule
  \multirow{6}{*}{\tiny Category} & AGG & 204 & 129 (63.2\%) & 163 (79.9\%) & 170 (83.3\%) & \textbf{34 (16.7\%)} \\
   & CHG & 76 & 14 (18.4\%) & 25 (32.9\%) & 33 (43.4\%) & \textbf{43 (56.6\%)} \\
   & EVT & 120 & 39 (32.5\%) & 42 (35.0\%) & 45 (37.5\%) & \textbf{75 (62.5\%)} \\
   & EXT & 236 & 92 (39.0\%) & 156 (66.1\%) & 182 (77.1\%) & \textbf{54 (22.9\%)} \\
   & ORD & 56 & 16 (28.6\%) & 23 (41.1\%) & 30 (53.6\%) & \textbf{26 (46.4\%)} \\
   & SPC & 60 & 18 (30.0\%) & 34 (56.7\%) & 40 (66.7\%) & \textbf{20 (33.3\%)} \\
  \midrule
   & \textbf{Total} & \textbf{752} & \textbf{308 (41.0\%)} & \textbf{443 (58.9\%)} & \textbf{500 (66.5\%)} & \textbf{252 (33.5\%)} \\
  \bottomrule
  \end{tabular}
\end{table}



\noindent\textbf{Quantify the Precision Loss of Approximate Queries.} Table~\ref{tab:gold_precision_loss_tolerance} quantifies how often the answers of the 752 approximate gold queries agree with those of their exact counterparts, as a function of the accepted error. With a tolerance of 1\%, only 41.0\% of the total answers agree (see bottom line of Table ~\ref{tab:gold_precision_loss_tolerance}). Relaxing the tolerance to 5\% and 10\% raises the agreement to 58.9\% and 66.5\%, respectively. The remaining 33.5\% failed all acceptance as an irreducible loss that no practical tolerance can recover. 

Of these 252 queries, 64 differ in the structural specification of their result, i.e., the two answers have a different number of rows or columns, or the approximate query returns nothing, while the other 188 differ by more than 10\% in at least one value. 

This loss grows steadily with semantic hardness, from 26.8\% for easy queries to 50.0\% for hard ones, and with execution hardness, from 20.0\% for small queries to 38.0\% for large ones. These results show that structurally complex queries over large regions are the most sensitive to the approximation. 

The query categories also have an impact on how much a looser tolerance helps. Agreement for extrema queries rises from 39.0\% to 77.1\%, whereas event queries barely benefit and only improve from 32.5\% to 37.5\%, with 31 of their 75 remaining disagreements in structural specification. 

Aggregation queries (\textbf{AGG}) are the least affected category with an irreducible loss of 16.7\%, while queries in events, thresholds and persistence (\textbf{EVT}) are affected the most, at 62.5\%, followed by queries in change, trend and comparison (\textbf{CHG}) at 56.6\%. 

This divergence motivates a closer look at why the same spatial expansion affects query types so differently. For instance, the aggregation question \textit{What was the average maximum wind speed in Canton of Bern during January 2005?} returns 2.114 m/s as the exact result and 2.107 m/s as an approximation (error=0.35\%). 
However, the picture changes notably in CHG: \textit{What was the precipitation anomaly in City of Zurich during 2010 compared to the 1981–2010 baseline?}, which yields +0.092 mm/day as the exact result but +0.541mm/day as an approximation -- a six-fold difference that shifts the interpretation from near-normal to clearly above-average. 

These two examples demonstrate that in some cases approximate results are sufficient while in other cases they might lead to insufficiently precise results. Note that data scientists working with STRATOS now have the option to choose between exact queries and approximate queries, where the latter trade precision for a lower runtime, in particular for large queries.


\subsection{RQ3: Ablation Study}

To justify the multi-agent architecture, we conducted an ablation study isolating the impact of key components within the STRATOS pipeline. This ablation is evaluated specifically on the agentic \textit{OpenSearch-SQL} framework backed by the \textit{Qwen} backbone model, as this setting  showed the best results in our previous experiments. The results for exact queries are summarized across both zero-shot and few-shot (FS) settings in Table~\ref{tab:ablation_qwen3}.

\begin{table}[htbp]
\centering
\caption{Ablation study for 752 exact queries on Qwen3 (OpenSearch-SQL) across zero-shot and few-shot (FS) settings. ST = Semantic Translation , SN = Symbolic-Numeric.} 
\label{tab:ablation_qwen3}
\resizebox{\linewidth}{!}{%
\begin{tabular}{ll cc c }
\toprule
\textbf{Ablation} & \textbf{Few-shot} & \textbf{EX Acc} & \textbf{VES} & \textbf{Mean Inf. Time (s)}  \\
\midrule
  \multirow{3}{*}{Full} & zero & 36.30\% & 0.360 & 74.4 \std{53.8}  \\
   & FS@1 & 43.48\% & 0.427 & 71.6 \std{51.6}  \\
   & FS@3 & \textbf{52.53\%} & \textbf{0.511} & 70.8 \std{52.9}  \\
\cmidrule(lr){1-5}
  \multirow{3}{*}{$-$ Query Rewriter} & zero & 35.24\% & 0.349 & 74.5 \std{54.0}  \\
   & FS@1 & 43.48\% & 0.428 & 72.9 \std{54.0}  \\
   & FS@3 & 49.60\% & 0.488 & 71.8 \std{52.9}  \\
\cmidrule(lr){1-5}
  \multirow{3}{*}{$-$ ST \& SN Bridging} & zero & 26.73\% & 0.263 & 80.7 \std{62.6}  \\
   & FS@1 & 33.51\% & 0.330 & 70.9 \std{51.3}  \\
   & FS@3 & 38.03\% & 0.373 & 71.0 \std{53.4}  \\
\cmidrule(lr){1-5}
  \multirow{3}{*}{$-$ Geo Context} & zero & 6.78\% & 0.062 & 64.9 \std{37.2}  \\
   & FS@1 & 9.31\% & 0.072 & 51.4 \std{34.7}  \\
   & FS@3 & 9.57\% & 0.065 & 45.2 \std{30.4}  \\
\bottomrule
\end{tabular}
}
  \rmspace
\end{table}


As observed in Table~\ref{tab:ablation_qwen3}, each module plays a distinct role in ensuring scalability and correctness. Removing the \textit{spatial query rewriter ($-$ Query Rewriter)} reduces EX from 52.53\% to 49.60\% for 3-shot queries (FS@3) and VES drops from 0.511 to 0.488. While inference time and baseline accuracy remain relatively stable, the database struggles with unoptimized queries, as evidenced by the total DB Runtime increasing from 581.8s to 786.9s for 1-shot queries (FS@1). Omitting the \textit{semantic translations ($-$ ST \& SN Bridging)} leads to measurable drops in both EX and VES dropping from 49.60\% to 38.03\% and from 0.488 to 0.373, respectively at FS@3 - reaffirming the necessity of the ontology mapping for accurately parsing complex meteorological concepts. Finally, removing the \textit{geospatial decoding component ($-$ Geo Context)} has the highest impact, forcing the LLM to hallucinate spatial coordinates. This causes EX to plummet dramatically from 38.03\% to 9.57\% at FS@3. 

The mean inference time in Table~\ref{tab:ablation_qwen3} covers the full end-to-end pipeline, including all LLM calls. It does not grow with the number of few-shot examples, because each example adds only about 240 tokens to a prompt of roughly 3.5k tokens. Latency is dominated by decoding about 2.7k output tokens per call, and the five candidate-generation calls run in parallel. At FS@3, the inference time is even slightly lower, largely because the five candidates agree more often. Fewer distinct queries have to be executed, aligned, and, if needed, corrected. 

\subsection{Error Analysis}

\begin{figure}[htbp]
  \centering
  \includegraphics[trim=25mm 20mm 25mm 30mm, clip, width=\linewidth]{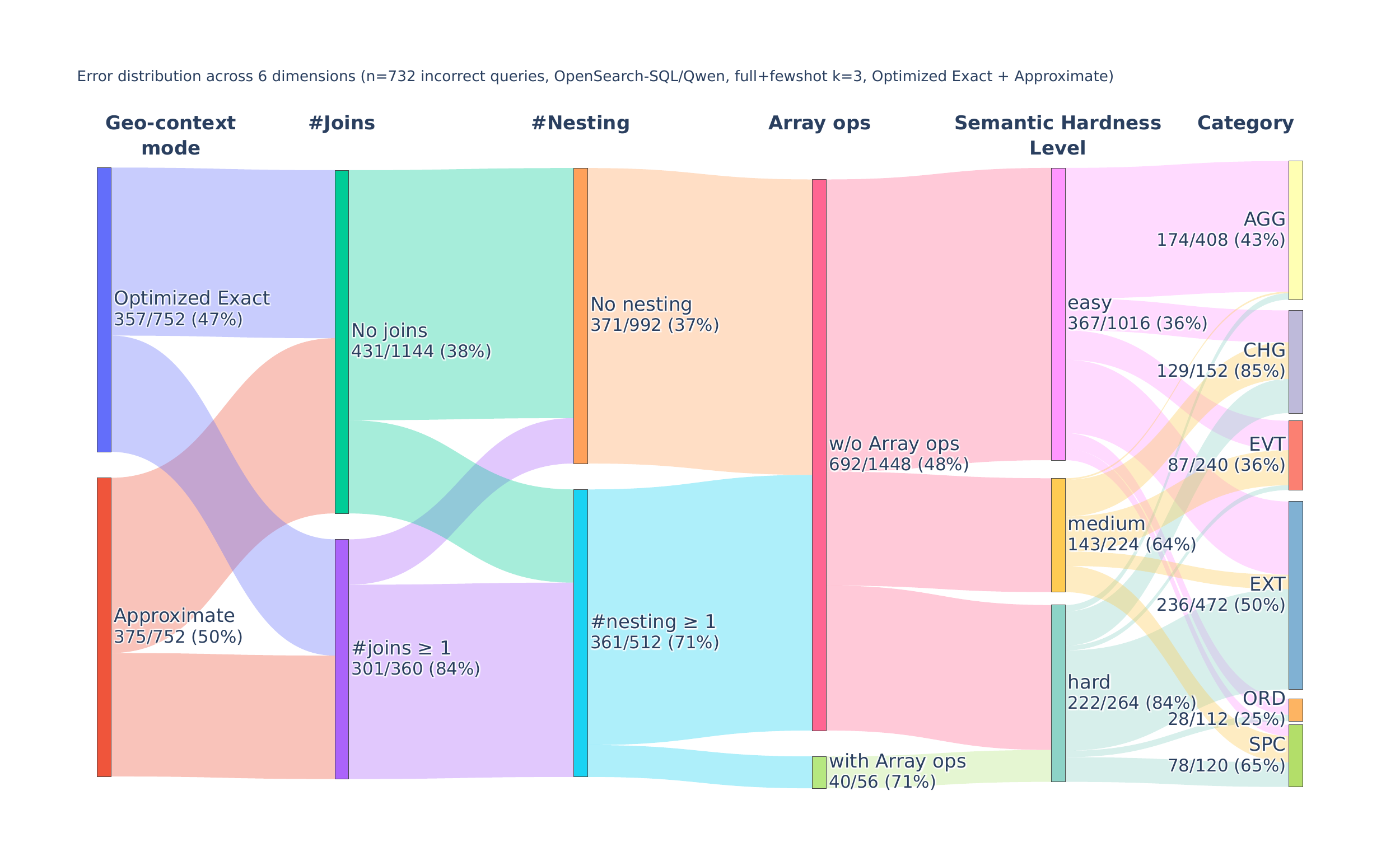}
  \caption{Error distribution across 6 dimensions with 732/1504 incorrect queries on OpenSearch-SQL/ Qwen3-Coder-Next (few-shot k=3, Optimized Exact + Approximate).}
  \label{fig:error-analysis}
\end{figure}

We perform error analysis on the best-performing configuration, \textbf{OpenSearch-SQL + STRATOS} with \emph{Qwen3-Coder-Next-80B} (few-shot $k{=}3$), across the 1,504 test instances spanning both geo-context modes (see Figure~\ref{fig:error-analysis}). The model fails on 48.7\% of queries overall (732/1,504), with a higher error rate under Approximate geo-context (50\%, 375/752) than under Optimized Exact (47\%, 357/752); this gap is consistent with the underlying predicate shift, since Optimized Exact resolves a place to a set of points following an \textsc{IN} operator, while Approximate needs to expand the same place into a bounding box requiring a four-sided range predicate (two \textsc{BETWEEN} operators on both ordered latitudes and longitudes). The latter raises the possibility that a syntactically plausible but boundary-incorrect SQL still passes generation, yet returns a mismatched result set. 

Structural complexity is the strongest driver of failure: queries requiring at least one join fail 84\% of the time versus 38\% for join-free queries. Queries with non-trivial nesting (CTEs/subqueries) fail 71\% versus 37\% without; queries using array operations fail at 71\%. Error rate also rises monotonically with our semantic-hardness tiers, from 36\% (easy) to 64\% (medium) to 84\% (hard), confirming the tiers track real difficulty. Breaking down by query category, Change/Comparison (CHG, 85\%) and Spatial Coverage (SPC, 65\%) queries are hardest, while Ranking/Distribution (ORD, 25\%) queries are comparatively easy, indicating that the model's weaknesses are concentrated in comparative reasoning and elevation/coverage-style spatial filters rather than simple aggregation or extrema lookups. 

\begin{table*}[htbp]
  \centering
  \caption{Primary causes of the 732 incorrect queries of OpenSearch-SQL + STRATOS (Qwen3-Coder-Next-80B, few-shot $k{=}3$), per geo-context mode.}
  \label{tab:error_causes}
  \footnotesize\setlength{\tabcolsep}{4pt}
  \begin{tabular}{@{}>{\raggedright\arraybackslash}p{1.45cm}>{\raggedright\arraybackslash}p{2.55cm}>{\raggedright\arraybackslash}p{\dimexpr\textwidth-1.45cm-2.55cm-0.8cm-1.05cm-1.5cm-10\tabcolsep\relax}>{\raggedleft\arraybackslash}p{0.8cm}>{\raggedleft\arraybackslash}p{1.05cm}>{\raggedleft\arraybackslash}p{1.5cm}@{}}
    \toprule
    \textbf{Family} & \textbf{Cause} & \textbf{Description} & \textbf{Exact} & \textbf{Approx.} & \textbf{Total (\%)} \\
    \midrule
    \multirow{3}{=}{Output Format} & Projection mismatch & \scriptsize{Correct values, but a different set or order of result columns (e.g., the gold query also returns the filter columns year and month), or an empty answer represented differently.} & 106 & 91 & 197 (26.9) \\
     & Granularity mismatch & \scriptsize{Result at a different aggregation level, e.g., one pooled value instead of one value per year for ``\textit{
     between 2005 and 2010}''.} & 43 & 47 & 90 (12.3) \\
     & Tied extrema & \scriptsize{Several grid cells share the minimum or maximum (e.g., a snow depth of 0), and the prediction returns a different one of them.} & 11 & 6 & 17 (2.3) \\
    \midrule
    \multirow{4}{=}{Spatial Reasoning} & Cross-resolution join & \scriptsize{The 0.1$^\circ$ meteorological grid is joined with the finer elevation or land-cover table on raw coordinate equality, which keeps only the cells with coinciding coordinates.} & 48 & 52 & 100 (13.7) \\
     & Geo-context neglect & \scriptsize{The provided place geometry is not used: a self-invented bounding box, another place (e.g., the \textit{City} instead of the \textit{Canton} of Bern), or no spatial filter at all.} & 44 & 13 & 57 (7.8) \\
     & Boundary shift & \scriptsize{Approximate only: the range predicate is written differently from the gold query (e.g., \texttt{BETWEEN} vs.\ \texttt{<}, rounded vs.\ raw coordinates), so boundary cells enter or leave the region.} & 0 & 32 & 32 (4.4) \\
     & Elevation-point selection & \scriptsize{``Highest/lowest point'' is resolved to a different elevation point than in the gold query.} & 6 & 6 & 12 (1.6) \\
    \midrule
    \multirow{5}{=}{Domain Logic} & Concept definition mismatch & \scriptsize{A meteorological concept or threshold is computed with a different definition (e.g., rainy day, heat wave, wet season).} & 37 & 43 & 80 (10.9) \\
     & Aggregation error & \scriptsize{Wrong or missing aggregate function (e.g., \texttt{AVG} instead of \texttt{SUM}).} & 10 & 11 & 21 (2.9) \\
     & Unit conversion error & \scriptsize{Missing or wrong unit conversion (e.g., \textit{Kelvin} vs.\ $^\circ C$, \textit{meters} vs.\ \textit{millimeters}).} & 10 & 15 & 25 (3.4) \\
     & Missing derived value & \scriptsize{A requested derived quantity is not computed, e.g., the change over time asked for by ``\textit{how has \dots\ evolved}''.} & 8 & 9 & 17 (2.3) \\
     & Temporal filter error & \scriptsize{Wrong time restriction (e.g., ISO week vs.\ calendar days, missing month filter).} & 7 & 6 & 13 (1.8) \\
    \midrule
    \multirow{2}{=}{Execution Failure} & Timeout & \scriptsize{The query exceeds the execution time limit.} & 25 & 28 & 53 (7.2) \\
     & Invalid SQL & \scriptsize{The query cannot be executed (e.g., \texttt{ROUND} on a double-precision value without a cast to \texttt{numeric}).} & 2 & 16 & 18 (2.5) \\
    \midrule
    \textbf{Total} & & & \textbf{357} & \textbf{375} & \textbf{732 (100)} \\
    \bottomrule
  \end{tabular}
\end{table*}

\noindent\textbf{Error Causes.} To understand \emph{why} queries fail, we assigned each of the 732 incorrect predictions its primary cause, i.e., the cause that makes its result differ from the gold result, using rule-based checks on the predicted SQL and its result followed by manual inspection. 

Table~\ref{tab:error_causes} groups the causes into four families: an \emph{Output Format} error returns the right content in a different form, a \emph{Spatial Reasoning} error selects the wrong grid cells or places, a \emph{Domain Logic} error computes a different meteorological quantity than asked, and an \emph{Execution Failure} returns no result at all. Output Format is the largest family (41.5\%). Its main cause, projection mismatch (26.9\%), is not a semantic error but arguably a partially correct answer: for instance, for ``Report the average precipitation in Attica on 2005-10-08'' the gold query returns the date columns next to the value, whereas the prediction returns the value only. Accepting such answers would raise EX from 52.5\% to 66.6\% under Optimized Exact and from 50.1\% to 62.2\% under Approximate geo context. 

Granularity mismatches (12.3\%) mostly stem from ambiguous questions: most of them ask for a statistic over a range of years, e.g., ``What was the average total precipitation in Canton of Bern between 2005 and 2010?'', which can mean one average per year (gold query) or a single average over the whole period (prediction). Both interpretations are valid, but only the former matches the gold result.

Spatial Reasoning accounts for 27.5\% of the errors. Its most frequent cause is the cross-resolution join (13.7\%), which silently keeps only the few cells whose coordinates intersect with the coordinates of both tables. For instance, for ``Show the average total precipitation (in mm) in Canton of Bern above 500\,m between 2015 and 2016'' the prediction joins the precipitation grid to the finer elevation table on raw coordinates (without rounding) and thus averages over a slightly different subset of cells, returning 4.523\,mm instead of 4.510\,mm. Note the the gold query rounds all coordinates to a 0.1 degree grid as shown below.

\begin{Verbatim}[commandchars=\\\{\},fontsize=\footnotesize,bgcolor=OliveGreen!15,breaklines]
\PYGZhy{}\PYGZhy{} Gold query: All coordinates (latitude and longitude) are rounded to a 0.1 degree grid.
ON ROUND(CAST(tp.latitude AS DECIMAL),1) = ROUND(CAST(e.latitude AS DECIMAL),1)
  AND ROUND(CAST(tp.longitude AS DECIMAL),1) = ROUND(CAST(e.longitude AS DECIMAL),1)
\end{Verbatim}
\begin{Verbatim}[commandchars=\\\{\},fontsize=\footnotesize,bgcolor=Salmon!20,breaklines]
\PYGZhy{}\PYGZhy{} Predicted query: Raw coordinates (without rounding) are used in the join, and only exactly intersecting cells are returned.
ON tp.latitude = e.latitude AND tp.longitude = e.longitude
\end{Verbatim}

Domain experts regard a part of these answers as partially correct as well. It is followed by geo-context neglect (7.8\%).

Domain Logic accounts for 21.3\%; more than half of these errors compute a meteorological concept with a different definition than the gold query, e.g., a rainy day defined by any single grid cell instead of the regional daily sum exceeding 1\,mm. 

Execution Failures are comparatively rare (9.7\%): timeouts are evenly split between both modes, whereas invalid SQL occurs almost only under Approximate geo context (16 vs.\ 2).

The causes also explain the gap between both geo-context modes. Of the 62 questions that fail only under Approximate geo context, 22 are boundary shifts: the model writes the range predicate in a different but plausible form, so that a few boundary cells enter or leave the region and the result changes slightly. Such errors cannot occur under Optimized Exact, where the place is given as an explicit set of grid cells.

A further class of discrepancies is unavoidable and stems from the dataset itself: ERA5-Land does not provide meteorological values for every grid cell, and data are missing in particular for some coastal cells. A bounding box, however, also covers cells outside the exact place geometry. For instance, the lowest elevation point within the bounding box of Attica (37.87$^\circ$N, 23.73$^\circ$E, $-4.66$\,m) lies on land near the coastline, but its 0.1$^\circ$ grid cell (37.9, 23.7) has no temperature records. Hence, the approximate gold query for ``the lowest temperature at the lowest point in Attica'' returns an empty result, whereas the exact query, restricted to the grid cells of the place itself, selects a different lowest point (37.99$^\circ$N, 23.40$^\circ$E) and returns a valid answer. Such errors are caused by missing data in ERA5-Land rather than by the Text-to-SQL system.

Finally, a manual review of the incorrect predictions by meteorological domain experts indicates that a substantial share of the wrongly predicted queries are acceptable in practice. In manual assessment, the Output Format errors and the Cross-resolution join errors are labeled as partially correct rather than as wrong. Under this more lenient notion of correctness, the acceptance rate of OpenSearch-SQL + STRATOS would rise from 51.3\% to approximately 78\% (80.2\% under Optimized Exact and 76.2\% under Approximate geo context). We report this as an indication of the gap between strict result-set equality and practical usefulness; all numbers reported elsewhere in this paper use the strict criterion. 




\section{Related Work}

Text-to-SQL has been a long-standing research topic in the database community \cite{affolter2019comparative}. Common techniques are relationship-aware attention for schema linking in RATSQL \cite{wang2020rat}, schema pruning in RESDSQL \cite{li2023resdsql} or graph-based schema question modeling in RASAT \cite{qi2022rasat}, and LGESQL \cite{cao2021lgesql}. Recent advances are largely driven by large language models (LLMs), with research focusing on improving query generation through prompt optimization \cite{pourreza2023din, gao2024text}, multi-agent decompositions \cite{pourreza2025chase, nooralahzadeh2025multi,xie2025opensearch, wang2025mac}, and multi-candidate generation with selection \cite{liu2026xiyan} or voting mechanisms \cite{sheng2025csc}. Additional gains come from supervised fine-tuning on domain-specific datasets \cite{li2024codes} and reinforcement learning with verifiable rewards \cite{toteja2025context}. Recent work incorporates planning, candidate selection, execution feedback, and agent-based reasoning \cite{chung2025long}. 

These approaches have been evaluated on several benchmarks such as BIRD \cite{li2023can}, ScienceBenchmark \cite{zhang2023sciencebenchmark}, FootballDB \cite{furst2025evaluating} and Spider 2.0 \cite{lei2025spider}.  However, none of these benchmarks addresses the ambiguity and scalability issues of spatio-temporal queries that are inherent in meteorological queries that cover data over time-span of 30+ years with spatial extensions of cities, regions, countries or even continents.

Prior spatial Text-to-SQL and NL2GIS systems ~\cite{Kazazi11122026,10824870} typically target vector geometries where continuous predicates (\texttt{ST\_Intersects}, \texttt{ST\_DWithin}, etc.) are the natural abstraction. In contrast, the meteorological workloads we consider operate on pre-aggregated scientific grids (ERA5-Land at \(0.1^\circ\)). ECMWF's own data services follow the same principle: for regular lat-lon products such as ERA5 and ERA5-Land, a regional extraction returns just the original grid points that lie inside the area, without interpolation~\cite{ecmwf2026areaextraction, ecmwf2026areaforum}. Continuous geometry overlay therefore neither improves precision nor removes the bottleneck that accounts for most of the execution cost on meteorological tables with about 180 million rows for each, namely the large number of grid points that the spatial filter has to match.

\section{Discussion and Conclusion}
\label{sec:conclusion}

In this paper, we presented STRATOS, an end-to-end Text-to-SQL framework designed specifically to tackle the scalability and semantic challenges of meteorological data science. By combining a multi-agent architecture for resolving symbolic-to-numeric gaps with a deterministic query rewriter for spatial arrays, our system drastically reduces both execution latency and the occurrence of database timeouts. 
STRATOS is not merely a theoretical framework evaluated in a sandbox; it has been actively developed and refined over the past 1.5 years as a core technological infrastructure within a large-scale European Union project involving over 13 consortium partners. The architecture presented in this work is continuously evolving, driven by real-world data management challenges and direct feedback from meteorologists and data scientists using the system daily. 

Future work will focus on extending our Domain Logic Injector to accommodate real-time streaming sensor data and optimizing spatial indexing strategies for distributed, cloud-native deployments.

\section*{Acknowledgment}
This work was supported by DataGEMS, funded by the European Union’s Horizon Europe Research and Innovation Programme, under grant agreement number 101188416.

\bibliographystyle{IEEEtran}
\bibliography{refs}

\end{document}